\documentclass{IEEEtran}
\usepackage{amsmath,amsfonts}
\usepackage{algorithmic}
\usepackage{algorithm}
\usepackage{array}
\IEEEoverridecommandlockouts
\usepackage[caption=false,font=normalsize,labelfont=sf,textfont=sf]{subfig}
\usepackage{textcomp}
\usepackage{stfloats}
\usepackage{url}
\usepackage{verbatim}
\usepackage{caption}
\usepackage[utf8]{inputenc} 
\usepackage{graphicx}
\usepackage{cite}
\usepackage[switch]{lineno}
\usepackage{amsmath,amssymb,amsfonts}
\usepackage{orcidlink}
\usepackage{comment}
\usepackage{algorithmic}
\usepackage{textcomp}
\usepackage{tabularx}
\usepackage{titlesec}
\usepackage{graphicx}
\usepackage{subcaption} 
\usepackage{threeparttable}
\usepackage{textcomp}
\usepackage{xcolor}

\usepackage{cite}
\usepackage{hyperref}

\hypersetup{
    colorlinks=true,
    citecolor=blue,
    linkcolor=blue,
    urlcolor=blue
}

\newcommand{\irv}[1]{{\color{black} #1}}
\newcommand{\imv}[1]{{\color{black} #1}}

\hypersetup{
    hidelinks,
    colorlinks=true,
    citecolor=blue,
    linkcolor=blue,
    urlcolor=blue
}
\def\BibTeX{{\rm B\kern-.05em{\sc i\kern-.025em b}\kern-.08emT\kern-.1667em\lower.7ex\hbox{E}\kern-.125emX}}

\begin{document}
\title{Package‑Embedded Coupled Inductor Arrays for High-Performance Computing Power Delivery}

\author{\IEEEauthorblockN{
Rami Rasheedi \orcidlink{0009-0009-8410-8011},~\IEEEmembership{Graduate Student Member,~IEEE}, Salma Abdelzaher \orcidlink{0009-0003-7242-5665},~\IEEEmembership{Graduate Student Member,~IEEE}, Inna Partin-Vaisband \orcidlink{0000-0002-6399-6672},~\IEEEmembership{Senior Member,~IEEE}}

}



\maketitle

\begin{abstract}
A novel power delivery framework, comprising a package-embedded inductor topology and an inductance-island methodology, is introduced to maximize both inductance and current densities in vertical power delivery (VPD). The framework leverages multiple multi-phase converters, a common strategy in high-performance computing systems, to enhance efficiency and scalability. The proposed topology employs an array of tightly coupled spiral square inductors sharing a common magnetic rod, serving multiple converters operating in the same conversion phase. The array is optimized to maximize coupling and minimize conversion losses, achieving superior inductance and current densities of 250~nH/mm${^2}$ and 10~A/mm${^2}$, respectively. At the system level, the inductance-island methodology partitions the power delivery network into multiple islands, each dedicated to a converter phase and supplying a portion of the load current, thereby enabling scalable and efficient distribution. 
To validate the framework, the inductor array is designed and simulated in ANSYS Maxwell 3D and Mechanical exhibiting an average quality factor of 23.6 and efficiency of 97.4\% at 2~A load current, 6~V input, and 10~MHz switching frequency. The inductor array netlist is extracted from ANSYS and co-designed in Cadence Virtuoso with a distributed dual-phase power conversion system, ensuring joint optimization of passive and active components. The co-designed converter achieves a significant efficiency gain of 5.65\% on average and up to 11.04\% at 40~A load over a similar converter with uncoupled inductors, demonstrating the practical benefits of the approach.
\end{abstract}


\section{Introduction}
\begin{figure}[]
\vspace{-10pt}
\centerline{\includegraphics[scale = 0.42]{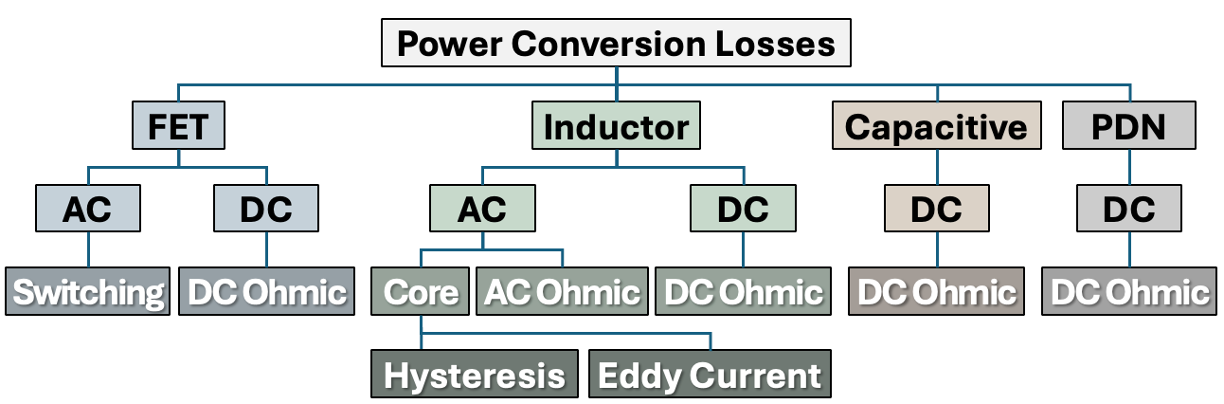}}
\caption{Power conversion loss components in high-power voltage regulators.}
\vspace{-10pt}
\label{system_loss}
\end{figure}

\IEEEPARstart{I}{n} modern high-performance computing (HPC), the main challenge for package-embedded high-step-down voltage conversion systems (e.g., 12~V to 1~V) is delivering high load current within a limited footprint while maximizing overall system efficiency~\cite{radhakrishnan2021power, 1}. This challenge arises from the combined contributions of multiple loss mechanisms, including those from power switches, inductors, capacitors, and the power distribution network, as illustrated in Fig.~\ref{system_loss}.

\begin{figure}[b!]
\centerline{\includegraphics[scale = 0.5]{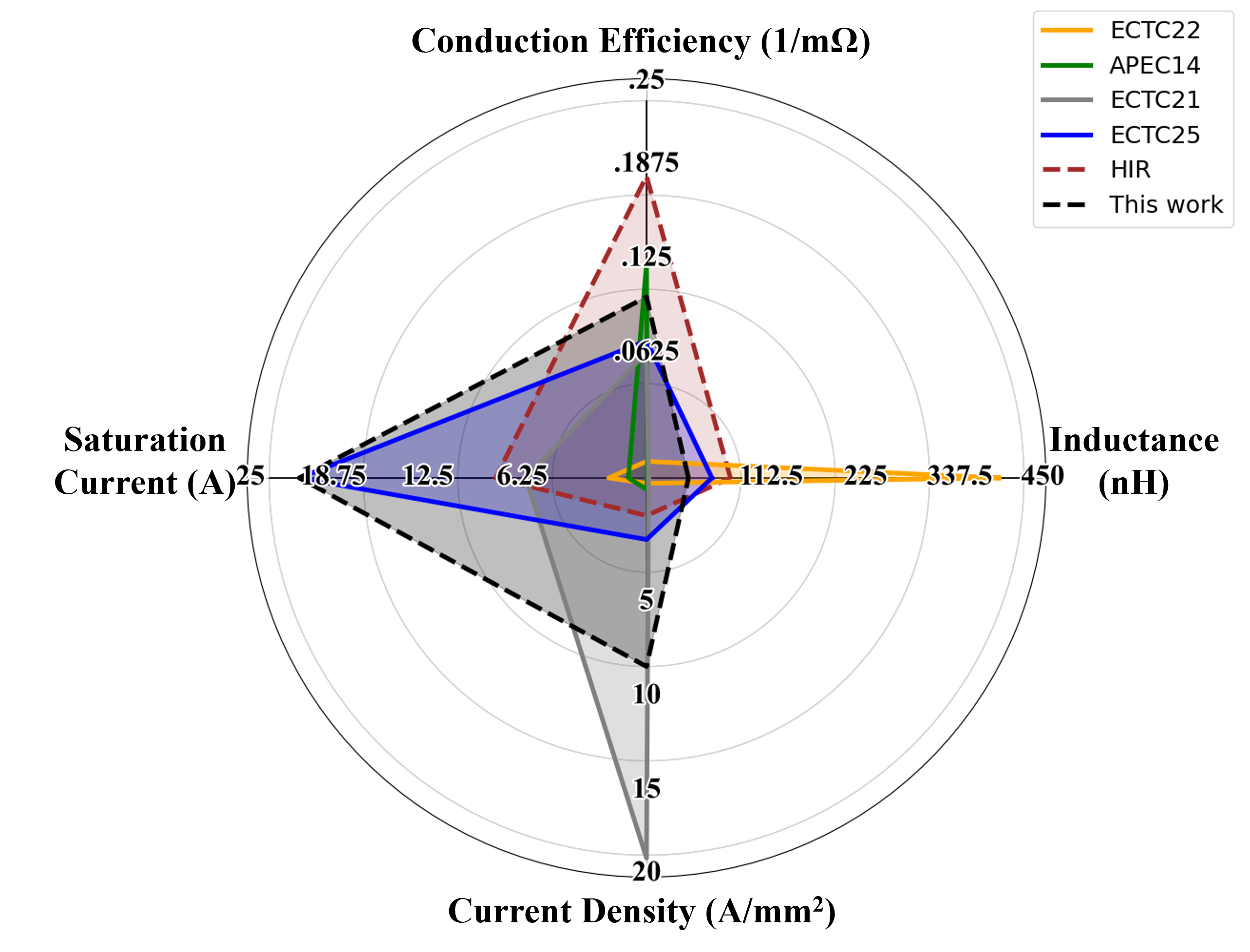}}
\caption{Comparison of embedded inductor design space across state-of-the-art, the proposed work, and HIR targets.}
\label{inductor_design_space}
\end{figure}

Inductors, key components in high-power conversion circuits, are subject to complex trade-offs among saturation current, inductance, size, DC resistance, and core losses. The performance of several state-of-the-art inductors~\cite{barros2021embedded, 9816521, avula2024design, burton2014fivr, sankarasubramanian2020magnetic, bharath2021integrated, ramiectc}, along with the target specifications outlined in the Heterogeneous Integration Roadmap (HIR)~\cite{ieeeHIR2023power, src-mapt-roadmap-2023}, is illustrated in Fig.~\ref{inductor_design_space}. While these inductors demonstrate excellent performance for certain metrics, none meet all the required specifications simultaneously. 

High inductance is a primary concern in HPC  systems and can be achieved through three main approaches: adding windings, increasing inductor size, or employing high-permeability cores. Each of these methods, however, incurs undesired trade-offs: (i) increasing the number of windings results in elevated DC resistance~\cite{8385217, 9124544}, (ii) utilizing a larger inductor incurs significant area overheads and also leads to higher DC resistance~\cite{9816521, 9097695}, (iii) high-permeability magnetic core materials may result in lower saturation currents owing to typically low saturation points in these materials~\cite{barros2021embedded}. 
Although the demand for high inductance decreases with increasing switching frequency, AC core losses in many high-permeability materials rise sharply at elevated frequencies~\cite{he2023soft}. As a result, the overall performance of power conversion systems is compromised when using existing inductors.

A promising approach for lowering both the inductance and saturation current requirements is multi-phase conversion~\cite{hinov2023design}.
A key advantage of the multi-phase approach is its inherent ability to mitigate total current ripple: the phase-shifted currents partially cancel at the output, allowing larger ripple in individual inductors without degrading voltage regulation. With higher allowable current ripple, the required inductance value can be reduced ($\Delta v = Ldi/dt$). Furthermore, distributing the DC output current across multiple inductors lowers the saturation current requirement for each individual inductor.

\begin{figure}[]
\centerline{\includegraphics[scale = 0.44]{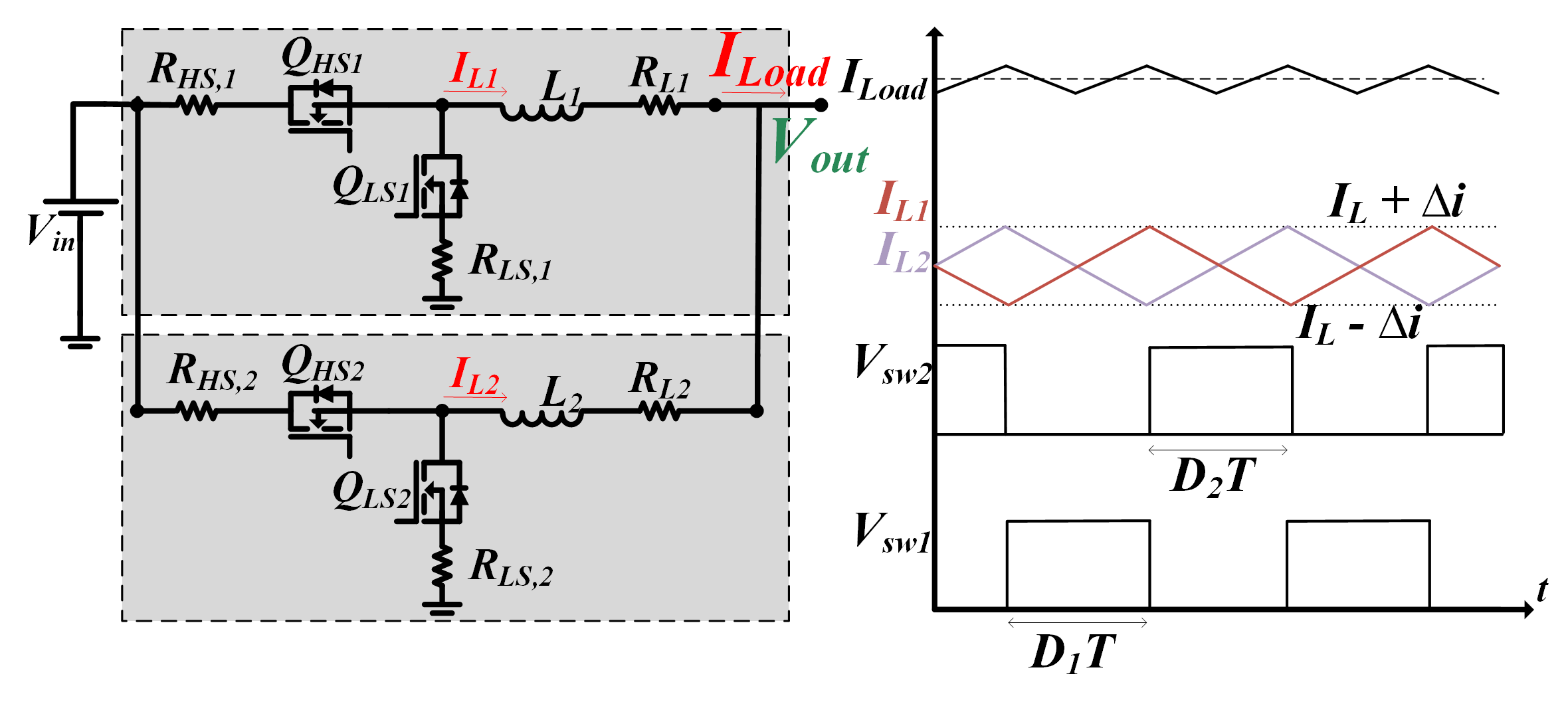}}
\caption{Circuit schematic of a two-phase buck converter with corresponding operating waveforms.}
\label{multiphase_Buck_converter}
\end{figure}

Current density is another primary concern in advanced power delivery systems. For example, modern AI systems may consume up to 1,000~W (e.g., 1,000~A at 1~V~\cite{10509290}) for dies as large as 250-800~mm$^{2}$. For scalability, the power delivery system should remain compact enough to fit within the footprint of the functional die (e.g., beneath the active die). This requirement places strict constraints on power inductors, which are typically the largest components in the system. Existing inductors are unable to simultaneously satisfy the required inductance and current density specifications. 

\imv{
Modern HPC systems require kW-level power delivery within stringent efficiency and thermal limits. This necessitates distributed vertical power delivery, where system-level current is supplied by many colocated VR units rather than by a single high-current converter~\cite{radhakrishnan2021power, 1}. Therefore, the key design objective is not maximizing current per VR, but achieving high current density and integration efficiency while enabling spatial replication across the package.
}

To address this challenge, a fundamentally novel inductor framework is proposed, consisting of a package-embedded inductor array co-designed with a new multi-phase conversion methodology for high-performance systems. With the proposed approach, the inductance density is enhanced through magnetic coupling, while current density is optimized to meet the stringent requirements of high-performance systems. 

\irv{Prior work on coupled inductors has primarily focused on coupling between different interleaved phases \cite{zhang2024organic, ding2020fan, wang2021novel, wong2001performance, li2022butterfly, elasser2023mini22, li2025air, elasser2023mini2}, where magnetic interactions alternate between positive and negative coupling depending on the phase relationship. As a result, the effective inductance fluctuates over the switching cycle, limiting both predictability and achievable inductance enhancement. To tune and increase the effective inductance, coupling between phase inductors and an auxiliary inductor has been proposed in \cite{zhu2022coupled}. However, this approach introduces additional area overhead and does not improve the current-handling capability of the power path.

In contrast, \textit{same-phase array-level} coupling is introduced in this work, resulting in a fundamentally different coupling regime. In this configuration, the effective inductance is substantially enhanced and stabilized without area or current-handling penalties. The array geometry (i.e., inductor dimensions, spacing, and layering) is analytically modeled, enabling efficient optimization of array density, effective inductance, and coupling under practical fabrication constraints. The operation of multiple inductors intentionally coupled within the same conversion phase is systematically analyzed and shown to produce a stable and significant increase in the effective inductance of each individual inductor with minimal impact on DC resistance, a regime that has not previously been addressed or quantified in the literature.
%
%
The primary contributions of this work are summarized as follows:
\begin{itemize}
    \item \textbf{A scalable tightly coupled inductor array architecture} for multi-phase, multi-converter power delivery systems. Unlike prior work focused on small sets of coupled inductors, the proposed architecture enables systematic scaling to large arrays suitable for high-current heterogeneous systems.
    
    \item \textbf{A system-level analytical model} describing the behavior of same-phase coupled inductor arrays in coordinated multi-phase and multi-converter operation, enabling co-optimization of ripple, efficiency, and scalability.
    
    \item \textbf{A practical co-design framework} integrating analytical modeling, electromagnetic simulation, and converter-level optimization to design large-scale coupled and uncoupled inductor arrays under realistic packaging and ripple constraints.
    
    \item \textbf{The first end-to-end design and evaluation} of a multi-phase, multi-converter system that systematically exploits tightly coupled same-phase inductor arrays, demonstrating performance regimes not achievable with conventional fixed or pairwise-coupled inductor designs.
\end{itemize}}

The remainder of this paper is organized as follows. Analytical models for power conversion and background on multi-phase converters are provided in Section~\ref{sec:background}.
The proposed inductor array topology and island-based power delivery methodology are presented in Section~\ref{Section_2}. The electrical and thermal performance of the inductor-array and overall power conversion system are presented in Section~\ref{Section_3} and  Section~\ref{Section_4}, respectively. The paper is concluded in Section~\ref{Conclusion}.

\section{Background}
\label{sec:background}

Hybrid voltage regulators, which integrate a switched-capacitor (SC) stage with a buck-type switching-mode power supply (SMPS), have emerged as a viable solution for power delivery in HPC systems~\cite{ECTC-salma, DPMIH44, DPMIH(N4K2), DSCB, DSCH, DPMIH62}. These two-stage architectures facilitate efficient step-down voltage conversion while supporting soft switching, thereby mitigating the hard-switching losses that are typically associated with the SC stage. Furthermore, the SC (first) stage relaxes the duty cycle constraints of the subsequent buck stage, thereby easing the constraint on the pulse signal generator. This configuration also lowers the inductance requirements for the buck inductor, as the voltage step applied to the second stage is reduced. For instance, in a 12 V to 1 V conversion, a 12~V input can be converted into an intermediate 6~V level by the SC stage, and the 6~V to 1~V conversion can be done in the buck stage. 



%

A two-phase approach with representative current waveforms is illustrated in Fig.~\ref{multiphase_Buck_converter} based on a buck converter architecture. 
The inductor current ripple can be approximated as
\begin{equation}
\Delta i \propto \frac{V_{in}-V_{out}}{L f_{sw}},
\label{eq:ripple_current_simple}
\end{equation}
showing an inverse dependence on inductance, $L$, and switching frequency, $f_{sw}$, and direct scaling with the voltage conversion step, $V_{in} - V_{out}$. 
Since ripple current contributes to AC ohmic losses and higher switching frequency elevates switching losses, an inductance–frequency trade-off must be optimized to constrain ripple within acceptable limits and maintain overall efficiency. 
Accordingly, the required inductance value should be selected with careful consideration of the targeted voltage step.

Although hybrid voltage regulation topologies relax inductor requirements, the inductors remain a dominant loss component of total power conversion losses. To provide a consistent analytical foundation for the reported efficiency and comparison with prior work, the individual loss mechanisms are summarized in Table~\ref{tab:inductor_loss}. These include DC and AC ohmic losses, magnetic hysteresis, and eddy current losses, which collectively determine the total inductor dissipation. The corresponding analytical expressions are employed in this work for loss estimation and validation of the proposed design methodology.
%
%

%
%

%
%
%
%

\begin{table}[]
\centering
\begin{threeparttable}
\caption{Power Loss versus Inductor Design Parameters$^*$.}
\label{tab:inductor_loss} 
\renewcommand{\arraystretch}{1.8}
\begin{tabularx}{\columnwidth}{c|X}
\textbf{Loss type} & \textbf{Expression} \\
\hline
DC ohmic& $\displaystyle I_{L,dc}^2  R_{L,dc}$ \\
AC ohmic& $\displaystyle \frac{2\Delta i^2}{D^2(1-D)^2} 
\sum_{k=1}^{N} \frac{\sin^2(k\pi D)}{(k\pi)^4} R_{L,ac}(2k\pi f_{sw})$ \\
AC hysteresis& $\displaystyle f_{sw} \oint H\, dB$ \\
AC Eddy current& $\displaystyle C B^2 f_{sw}^2 d^2 / \rho$ \\
\hline
\end{tabularx}
\begin{tablenotes}
\footnotesize
\item[*]\(R_{L,dc}\): winding DC resistance, \(I_{L,dc}\): inductor DC current,  \(\Delta i\): current ripple amplitude, \(D\): duty cycle, \(R_{L,ac}\): winding AC resistance, \irv{\(k\): harmonic index of the inductor current ripple (positive integer)}, \(f_{sw}\): switching frequency, \(H\): magnetic field intensity, \(B\): magnetic flux density, \( C \): proportional constant, \irv{\( d \)}: core thickness, \( \rho \): core material resistivity.
\end{tablenotes}
\end{threeparttable}
\end{table}

\section{Power Island-Aware Coupled Inductor Array}
\label{Section_2}
A central element of the proposed power delivery system is the coupled inductor array. 
The key idea is to employ vertically aligned inductors with a shared magnetic core, simultaneously increasing power efficiency and density. Particular attention is given to the choice of magnetic material, as its characteristics strongly influence performance metrics such as saturation current, permeability stability, and core loss. 

\begin{figure}[t]
\centerline{\includegraphics[scale = 0.38]{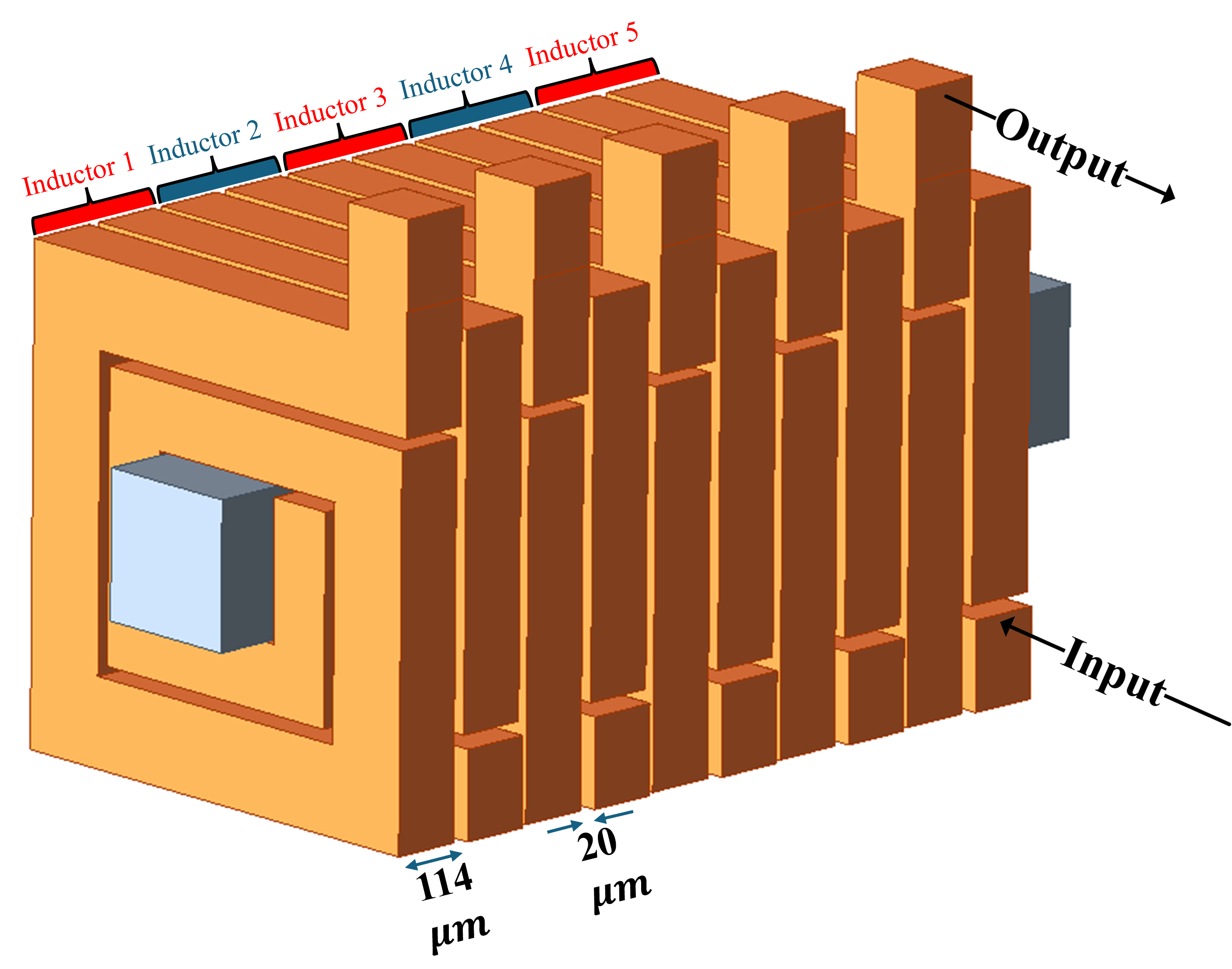}}
\caption{A five-inductor section of the proposed inductor array.}
\label{5inds_with_core_pins}
\end{figure}

\subsection{Array-Based Architecture}
The proposed architecture comprises an array of aligned, vertically integrated two-turn two-layer square inductors with shared magnetic core and input and output pins placed at the bottom and top of the structure, respectively. A five-inductor section of the array is shown in Fig.~\ref{5inds_with_core_pins}. 
The individual inductors are designed with variable-turn widths~\cite{ramiectc} to maximize the array quality factor. The height of the structure is limited to 800~\textmu m to accommodate fabrication and integration within a typical package substrate \cite{HIR_WLP_2021}. The number of turns is restricted to two to avoid excessive resistive losses within these height constraints. 

Enhancements in both inductance and current density can be achieved by reducing the thickness of the inductor layers and/or the separation between the metal traces. However, a lower layer thickness increases ohmic losses, and smaller trace separation is constrained by fabrication capabilities. In this work, trace separation is considered as a design constraint, and metal thickness is optimized for quality factor (L/R). 

\begin{figure}[]
\centerline{\includegraphics[scale = 0.4]{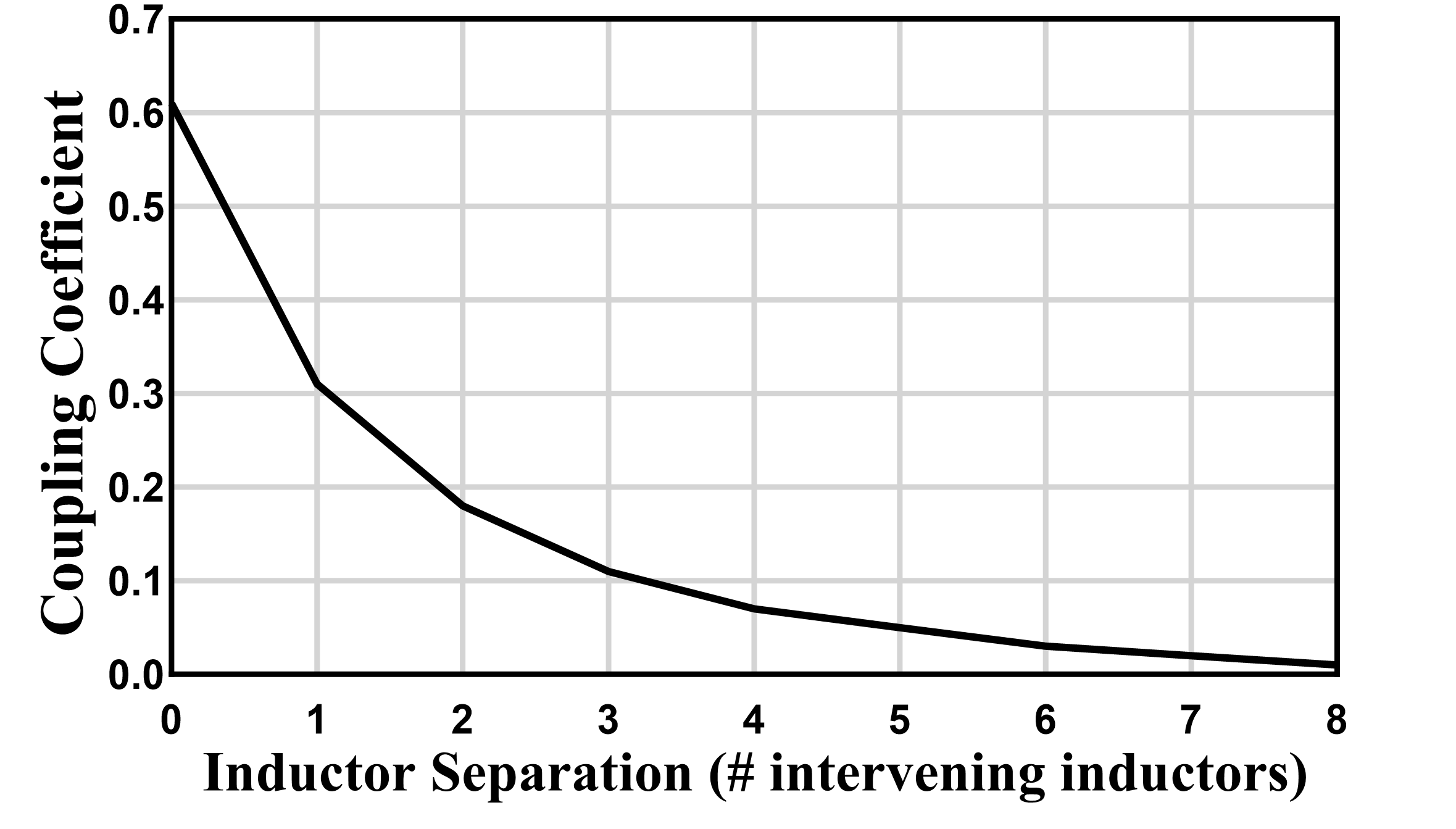}}
\caption{Coupling coefficients between two inductors separated by different numbers of inductors (aka inductor separation) in a 10-inductor array.}
\vspace{-10pt}
\label{coeff_Ls}
\end{figure}

Alternatively, wider inductors exhibit lower DC resistance at the expense of larger inductor size. The width should therefore be optimized to maximize the quality factor under the height constraints. Note that the height is constrained by both the available substrate thickness and the feasibility of fabrication processes. While the coupled inductance typically increases with the increasing number of inductors within the array, the coupling factor ultimately saturates due to negligible coupling between farther placed inductors, as shown in Fig.~\ref{coeff_Ls}. In a 10-inductor array, coupling saturates at inductor separation of six. This indicates that increasing the array beyond ten inductors does not substantially increase the peak inductance of a single inductor (since the peak coupling coefficient is already reached by closely spaced pairs). 
However, larger arrays increase the fraction of inductors operating near this maximum. For example, in a 15-inductor array, only the few central inductors exhibit strong coupling, whereas in a 25-inductor array, more than half of the inductors (those located away from the array edges) approach the maximum inductance. Thus, larger arrays improve inductance uniformity, current sharing, and scalability. \irv{Since Fig.~\ref{coeff_Ls} shows that coupling becomes negligible beyond a finite separation distance, inter-array coupling is neglected.}

\begin{figure*}[b]
    \centering
    \vspace{-20pt}
    
    \subfloat[]{\includegraphics[width=0.33\textwidth]{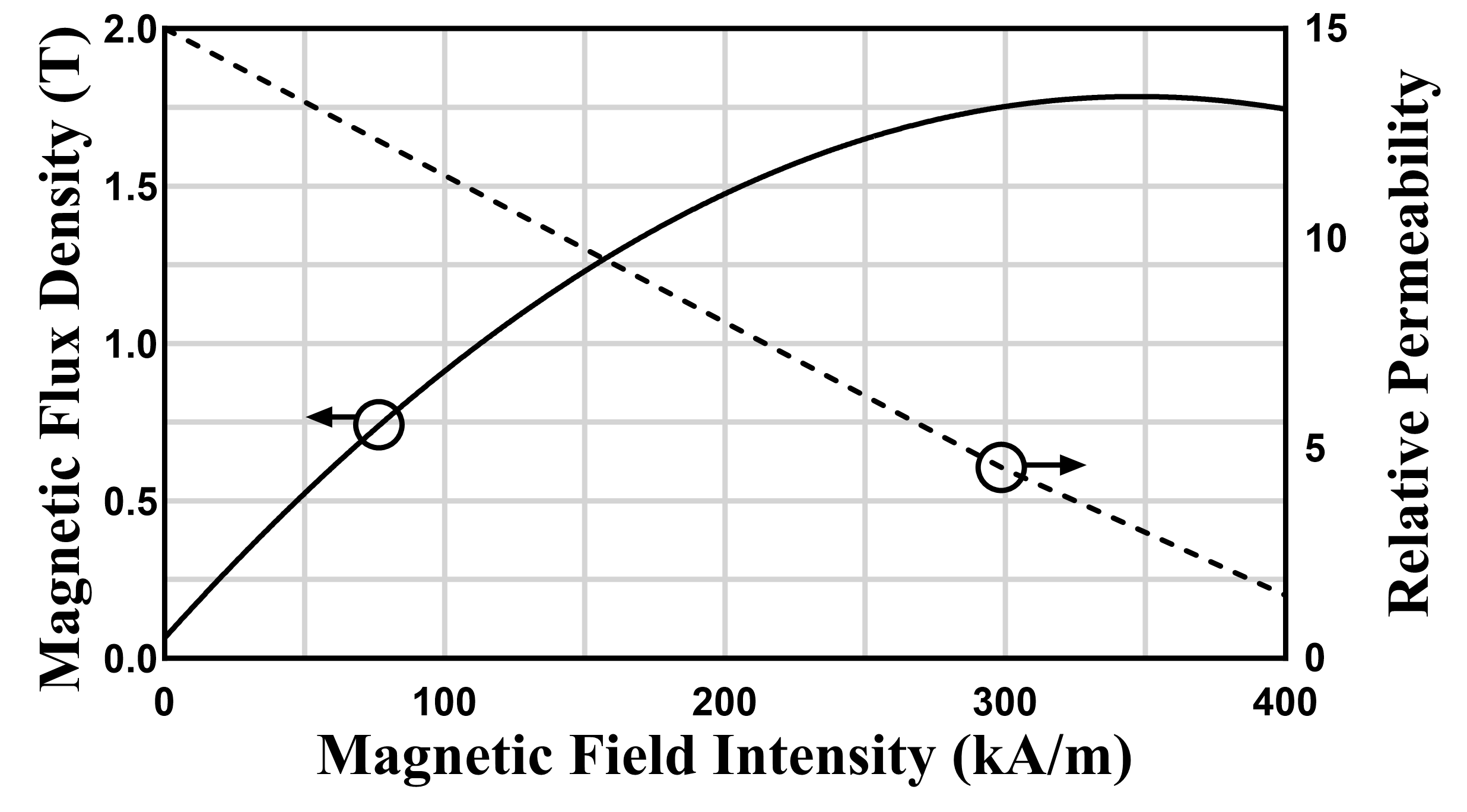}%
    \label{bh_curve_and_mue}}    
    \subfloat[]{\includegraphics[width=0.3\textwidth]{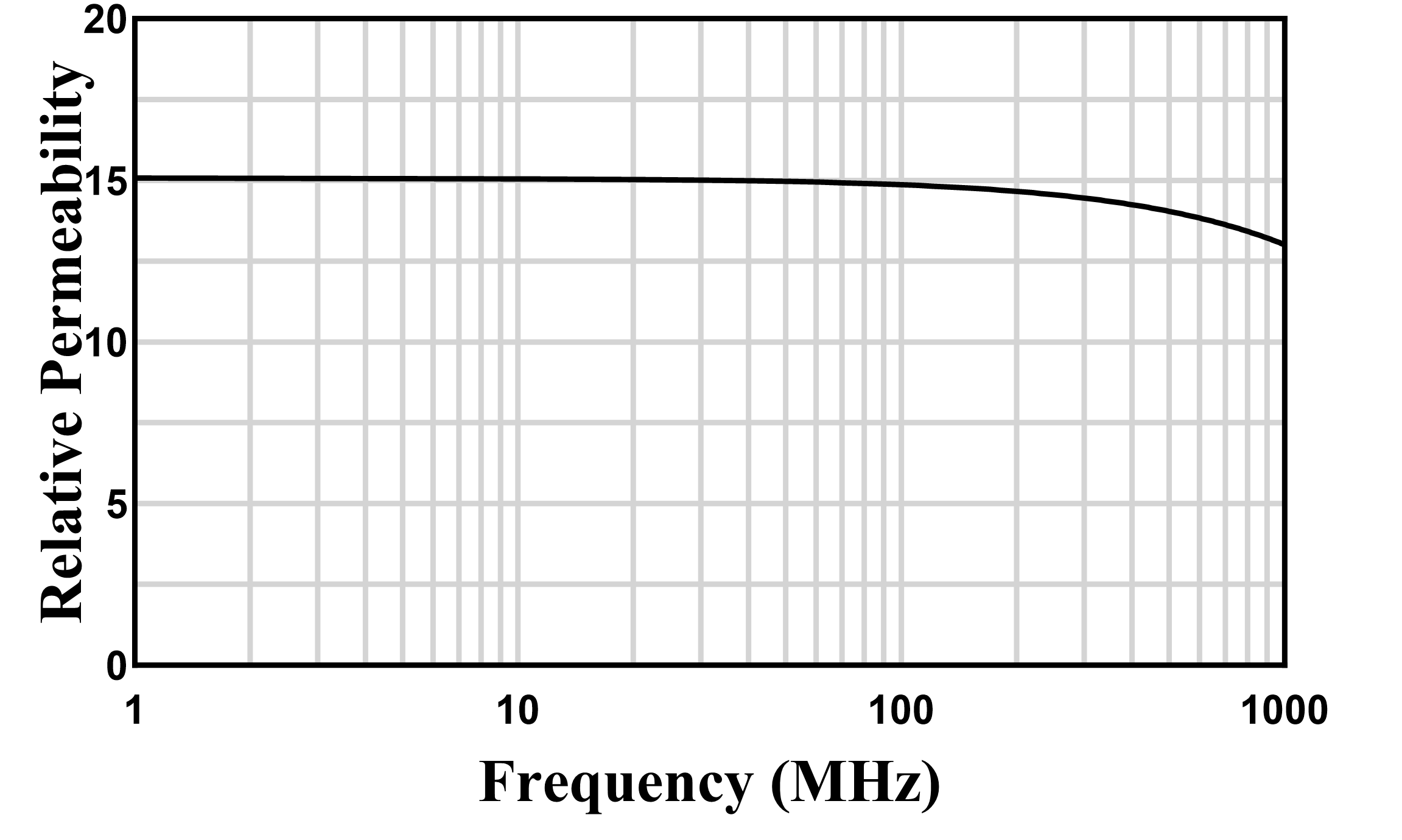}%
    \label{mue_curve}}    
    \subfloat[]{\includegraphics[width=0.32\textwidth]{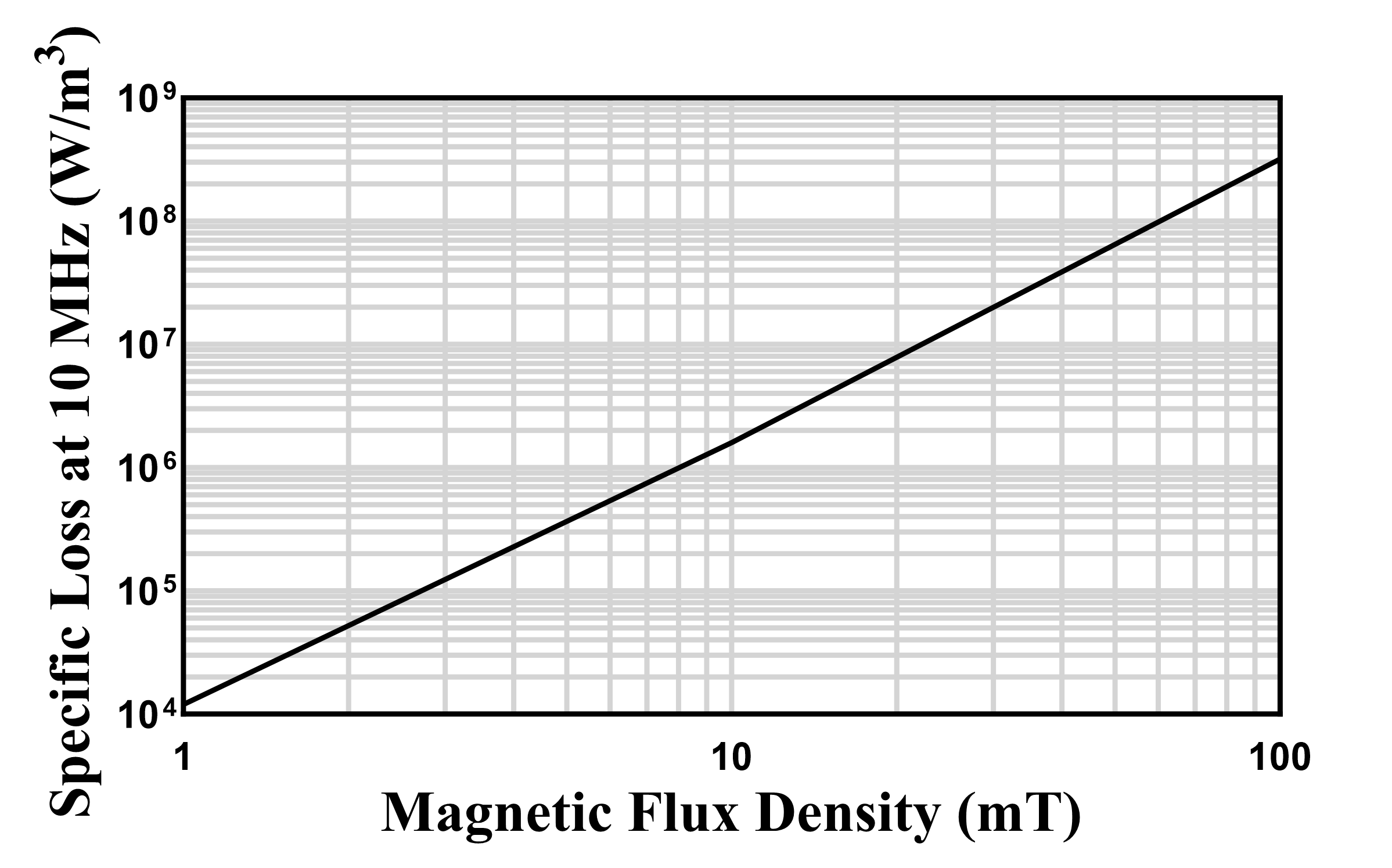}%
    \label{core_loss_curve}}
    
    \caption{Magnetic characteristics of TY-M5: (a) B-H relation and relative permeability, (b) relative permeability vs. frequency, and (c) core loss at 10 MHz vs. magnetic flux density.}
    \label{fig:TYM5_characteristics}
\end{figure*}


\subsection{Magnetic Core}
Magnetic cores are commonly used in embedded inductors to increase inductance density, but their usefulness is limited by the tradeoff among relative permeability, saturation point, resistivity, and high-frequency core loss.
Ferrite materials such as nickel-zinc (NiZn) and manganese-zinc (MnZn) provide useful permeability, but are often limited by saturation and insufficiently characterized loss behavior in the 5-100~MHz range relevant to embedded power delivery. 
Iron-based materials offer high permeability but exhibit low resistivity and consequently high AC core loss at these frequencies.

In this work, TY-M5 (developed by Taiyo Yuden~\cite{apec2025}) is selected because it combines high saturation flux density, stable permeability up to 100 MHz, and low core loss under the targeted operating conditions, as shown in  Fig.~\ref{fig:TYM5_characteristics} and summarized in Table~\ref{tab:Magnetic_materials_comparison}~\cite{he2023soft}. 

These results indicate that TY-M5 enables reduced frequency-dependent variations and achieves competitive loss performance, making it a promising material for high-frequency integrated inductors.

\imv{The proposed inductor array employs a shared magnetic rod, resulting in a partially open magnetic path. While closed magnetic loops improve flux confinement and inductance, they incur a significant footprint penalty, which is contrary to the objective of maximizing current density in area-constrained packages. The design, therefore, reflects a trade-off between magnetic confinement and integration density.
Magnetic losses are simulated using the frequency-dependent characteristics of the TY-M5 material (Fig.~\ref{core_loss_curve}) provided by the vendor, including both hysteresis and eddy-current losses within the core. These losses remain a small fraction of the total power dissipation under the targeted operating conditions.
The impact of the open magnetic path on flux leakage, stray coupling, and associated loss mechanisms is captured through full-wave electromagnetic simulations, which account for fringing fields and induced currents in surrounding conductors. The results indicate that these effects are limited. Based on this analysis, a separation distance beyond which magnetic coupling becomes negligible is identified, and inductor arrays associated with different phases are placed beyond this threshold to minimize coupling and EMI. The high current density further enables flexible placement without requiring close proximity.}



\begin{table}[]
\centering
\begin{threeparttable}
\caption{Magnetic Materials.}
\label{tab:Magnetic_materials_comparison} 
\begin{tabular}{c|l|l|l}
\textbf{Material} & \textbf{\textit{$B_{sat}$(T)}} & \textbf{\textit{$\mu_{r}(\times 10^{3})$}} & \textbf{$\rho(\Omega \cdot m)$}  \\
\hline
TY-M5 & 1.25 & 0.015 & N/A\tnote{*} \\
MnZn ferrites &0.4-0.55 &1-10 &$10^{-2}$ \\
NiZn ferrites & 0.2-0.5&0.1-1 &$10^{5}$ \\
Pure Fe&2.16&3-50& 10.5$\times 10^{-8}$\\
Oriented Fe$_{97}$Si$_{3}$ (wt.\%) &2.02  &15-80 &45$\times 10^{-8}$ \\
Fe$_{15}$Ni$_{80}$Mo$_{5}$ (wt.\%) &0.8  &500 &70$\times 10^{-8}$ \\
Fe$_{52}$Ni$_{48}$ (wt.\%) &1.6  &100 &48$\times 10^{-8}$ \\
Fe$_{49}$Co$_{49}$V$_{2}$ (wt.\%) &2.35  &2 &40$\times10^{-8}$ \\
Fe$_{78}$B$_{13}$Si$_{9}$ (wt.\%) &1.56  &100 & 120-140$\times10^{-8}$  \\
\hline
\end{tabular}
\begin{tablenotes}
\footnotesize
\item[*]Core loss data for TY-M5 is included
\end{tablenotes}
\end{threeparttable}
\vspace{-10pt}
\end{table}

\begin{figure}[b!]
\vspace{-15pt}
\centerline{\includegraphics[scale = 0.38]{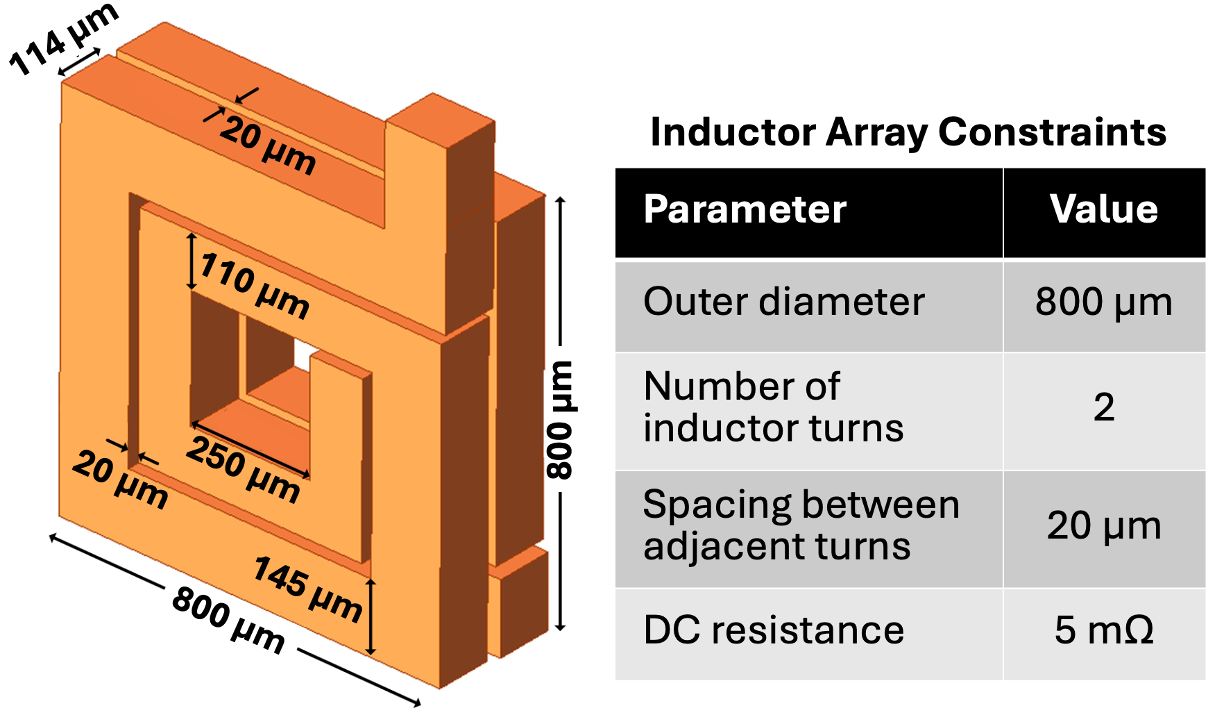}}
\caption{Individual inductor dimensions after optimization.}
\label{inductor_dims}
\end{figure}

\subsection{Array Optimization}
The approach proposed in~\cite{rasheedi2024embedded, ramiectc} is followed in the optimization of the individual inductors within the array, where each inductor is represented as a set of metal segments of uniform thickness with segment width varying according to the turn number. The analytical models are summarized in Table~\ref{tab:Physics-based_analytical_model_equations}. 
The optimization figure of merit (FOM) is defined as a function of inductance, $L$, DC resistance (DCR), and footprint as
\begin{equation}
\text{FOM} = \frac{L}{\text{footprint} \cdot \text{DCR}}.
\end{equation}
The FOM effectively balances DC loss (through the DC resistance), current quality and AC loss (through inductance), and current density (through the footprint). Alternatively, aspects related to the magnetic core (e.g., saturation current) are not captured by the FOM, which is focused on optimizing the inductor geometry. The spacing between individual inductors in the array is set to 20~\textmu m, a value commonly reported as the practical minimum \cite{avula2024design}, to ensure strong coupling. The physical constraints used for the optimization and optimized dimensions of a single inductor are shown in Fig.~\ref{inductor_dims}, where footprint is given by the product of the inductor thickness (248~\textmu m) and outer diameter (800~\textmu m)
reflecting the vertical orientation of the inductor. 

\begin{table*}[]
\centering
\begin{threeparttable}
\caption{Physics-Based Analytical Models for a Single Inductor in the Proposed Inductor Array as a Function of the Inductor Design Parameter$^*$}.
\label{tab:Physics-based_analytical_model_equations} 
\renewcommand{\arraystretch}{1.8}
\begin{tabularx}{\textwidth}{c|X}
\textbf{Quantity} & \textbf{Expression} \\
\hline
Width of the $k$\textsuperscript{th} segment& $W_k^{seg} = ak^b$\\
\hline
Average width of all segments& $W_{avg} = \frac{1}{N} \sum_{n=1}^{N} W_n$ \\
\hline
Total length of the inductor metal& $l_{total} = (4N+1) D_{in} + (4N+1)N(W_{avg} + S)$ \\
\hline
Average length of the segment& $l_{avg}^{seg} = \frac{l_{total}}{4N}$ \\
\hline
Self inductance of the inductor& $L_{self} = \sum_{k=1}^{4N} \frac{\mu}{2\pi} l_k^{seg} \left( \ln \frac{2l_k^{seg}}{W_{avg} + T} + 0.5 \right)$\\
\hline
Grover's mutual inductance formula~\cite{grover1946inductance}& $M= 
\frac{\mu}{2\pi} l \left( \ln \left( \text{ldr} + \sqrt{1 + \text{ldr}^2} \right) - \sqrt{1 + \frac{1}{\text{ldr}^2}} + \frac{1}{\text{ldr}} \right),~\text{ldr}=\frac{l}{d}$ \\
\hline
Negative mutual inductance of the inductor& $M^{-} = 0.47 \frac{\mu}{2\pi} N l_{total}$ \\
\hline
Average distance between same-side segments& $D_{mean}^{M^{+}} = \frac{1}{3}(W_{avg} + S)(N+1)$ \\
\hline
Positive mutual inductance of the inductor& $M^{+} =~\frac{\mu}{2\pi} l_{total}(N-1)
    \cdot\Big[ \ln \left(\text{LDR} + \sqrt{1 + \text{LDR}^2} \right)- \sqrt{1 + \frac{1}{\text{LDR}^2}} + \frac{1}{\text{LDR}} \Big],~\text{LDR}=l_{avg}^{seg}/D_{mean}^{M^{+}}$ \\
    \hline
Total inductance& $L = L_{self} + M^{+} - |M^{-}|$ \\
\hline
DC resistance of the inductor& $R_L = \frac{\rho}{T} \sum_{k=1}^{4N} \left(l_k^{seg}/W_k^{seg}\right)$ \\
\hline
\end{tabularx}
\begin{tablenotes}
\footnotesize
\item[*]$a$, $b$ are optimization parameters generated from the model to achieve the optimal turn-varying strategy that maximizes the FOM, $N$ is the number of inductor turns, $S$ is the inductor turn separation, $D_{in}$ is the inductor inner diameter, and $T$ is the inductor thickness
\end{tablenotes}
\end{threeparttable}
\end{table*}

\subsection{Island-Based Power Delivery}
\begin{figure}[]
\centerline{\includegraphics[scale = 0.45]{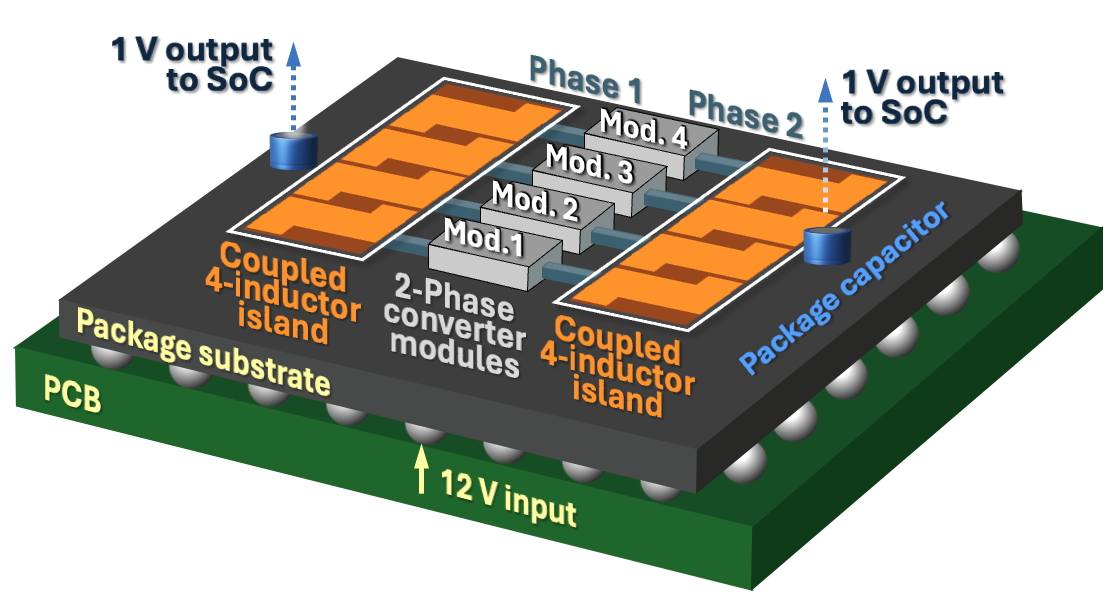}}
\caption{Island-based multi-phase power delivery scheme utilizing the proposed inductor arrays embedded in the package substrate.}
\vspace{-15pt}
\label{Inductance_islands_pp}
\end{figure}

An island-based multi-phase power conversion methodology is proposed to complement the coupled inductor array topology and enable scalable high-current distribution.
In this scheme, the proposed coupled inductors are tailored for multi-phase operation, with the overall inductor arrays systematically organized into units, or islands, that enable scalable and efficient power delivery. Each island corresponds to a distinct conversion phase within a power delivery system comprising multiple multi-phase converters, as depicted in Fig.~\ref{Inductance_islands_pp}. 
Such phase-aware design is critical for achieving simultaneous switching and thereby maximizing positive magnetic coupling. The number of inductors in each array is determined by the total number of converters in the power delivery system. 
This relationship depends on the system’s load requirements and on the current-supplying capability of each converter, which is primarily governed by the resistance, saturation, and thermal characteristics of the inductors. Furthermore, the number of arrays corresponds to the number of phases in the multi-phase power delivery system. For instance, a system that requires a current of 200~A, utilizing dual-phase converters with an inductor current of 2~A, will require four~25-inductor arrays. Each pair of arrays serves 25 converters, where the single array in each pair serves a converter phase, resulting in a total of two~25-inductor array pairs and 50 converters. 

Enhanced coupling among array inductors can increase the effective inductance of each inductor by up to 300\%. This improvement reduces reliance on high-permeability core materials, enabling the use of moderately permeable, low-loss, high-saturation magnetic materials. Such a choice also improves the trade-off between performance and material costs. In addition, this configuration supports a shared-inductor strategy in which several distributed converters operate in parallel, while utilizing a common inductor array within each phase. 

\section{Inductor Simulation Results}
\label{Section_3}
The proposed framework is demonstrated using a ten-inductor array. 
The electrical and thermal performance of the optimized array is evaluated through finite-element simulations. Key parameters, including self- and coupled inductance, resistance, and core loss over a broad frequency range, are extracted using electromagnetic modeling in ANSYS Maxwell 3D. 
These simulations provide quantitative insight into the impact of magnetic coupling, conductor geometry, and core material on the overall power delivery efficiency and reliability. When benchmarked against state-of-the-art inductors, coupled inductor arrays exhibit better performance in terms of inductance density, saturation current, and conduction loss.

\subsection{Electrical Analysis}

To evaluate the impact of magnetic coupling, the inductance of a standalone optimized two-turn inductor is compared with the average inductance of a two-turn segment within the proposed array, as shown in Fig.~\ref{L_data}. The coupled segment exhibits nearly four times higher inductance up to 100~MHz due to the shared magnetic core and close proximity of the inductors. Such enhancement is achieved when all inductors within the array switch simultaneously.   

To evaluate array power efficiency, both AC resistance and core losses are extracted as functions of frequency and flux density. The winding resistance up to 100~MHz is shown in Fig.~\ref{ACR_data}. The DC resistance (DCR) is 8.2~m\( \Omega \), which establishes the conduction-loss floor. With increasing frequency, the effective resistance rises due to skin and proximity effects. An average quality factor of approximately 23.6-44.4 is observed at 10-50 MHz.

Core losses are extracted at a switching frequency of 10~MHz with a 6~V voltage excitation amplitude. Based on the results (see Fig.~\ref{array_core_loss}), the instantaneous core loss peaks at 34.62~mW and averages 17.17~mW, comprising 0.08\% out of a 2~W per-inductor total power as targeted in this work. 

The inductance-current characteristic is shown in Fig.~\ref{Isat}, indicating a saturation current of $\sim$23~A. The inductance remains nearly constant up to 7~A, ensuring stable ripple behavior in this region. 
\imv{Although the inductor exhibits a high saturation capability, the operating current in this work is limited to 2~A per inductor based on thermal constraints discussed in the following subsection.}

An average inductor efficiency of \imv{97.4\%} is achieved at 2~A load, 6~V input, 10~MHz based on
\begin{equation}
\eta = \frac{I_{Load}V_{out}}{I_{Load}V_{out}+ P_{dc} + P_{ac}},
\end{equation}
with a per-inductor dc loss of 32~mW, an ac ohmic loss of 0.143~\textmu W, and a core loss of 1.71~mW at 10~MHz. 
To evaluate the efficiency of the converter, the inductor array netlist is extracted in ANSYS Maxwell 3D and simulated with Cadence Virtuoso, as described in Section~\ref{Section_4}.

{\captionsetup[subfigure]{font=footnotesize}
\begin{figure}[t]
\centering
\subfloat[]
{\includegraphics[width=0.24\textwidth]{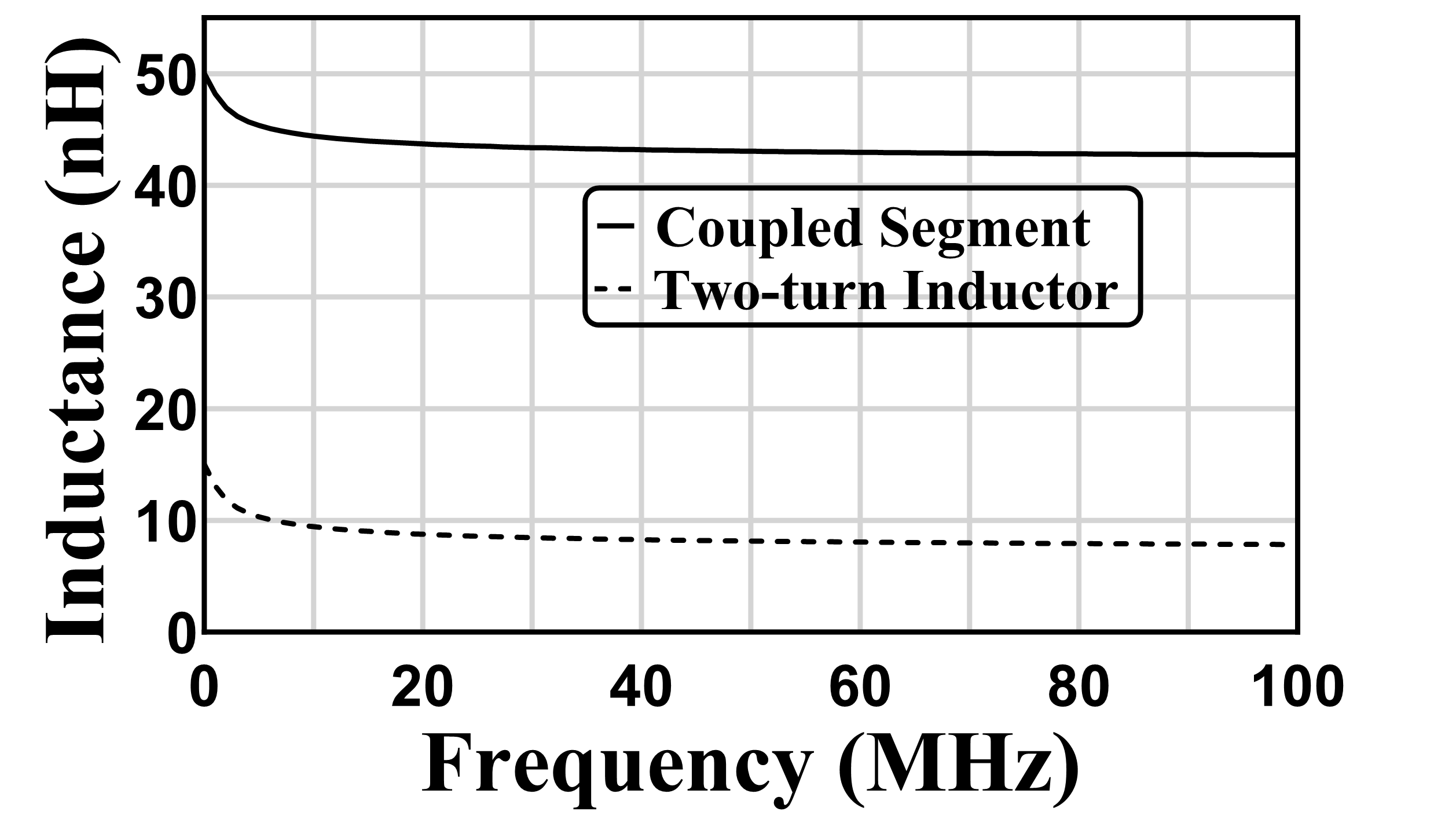}%
\label{L_data}}
\hfil
\subfloat[]{\includegraphics[width=0.24\textwidth]{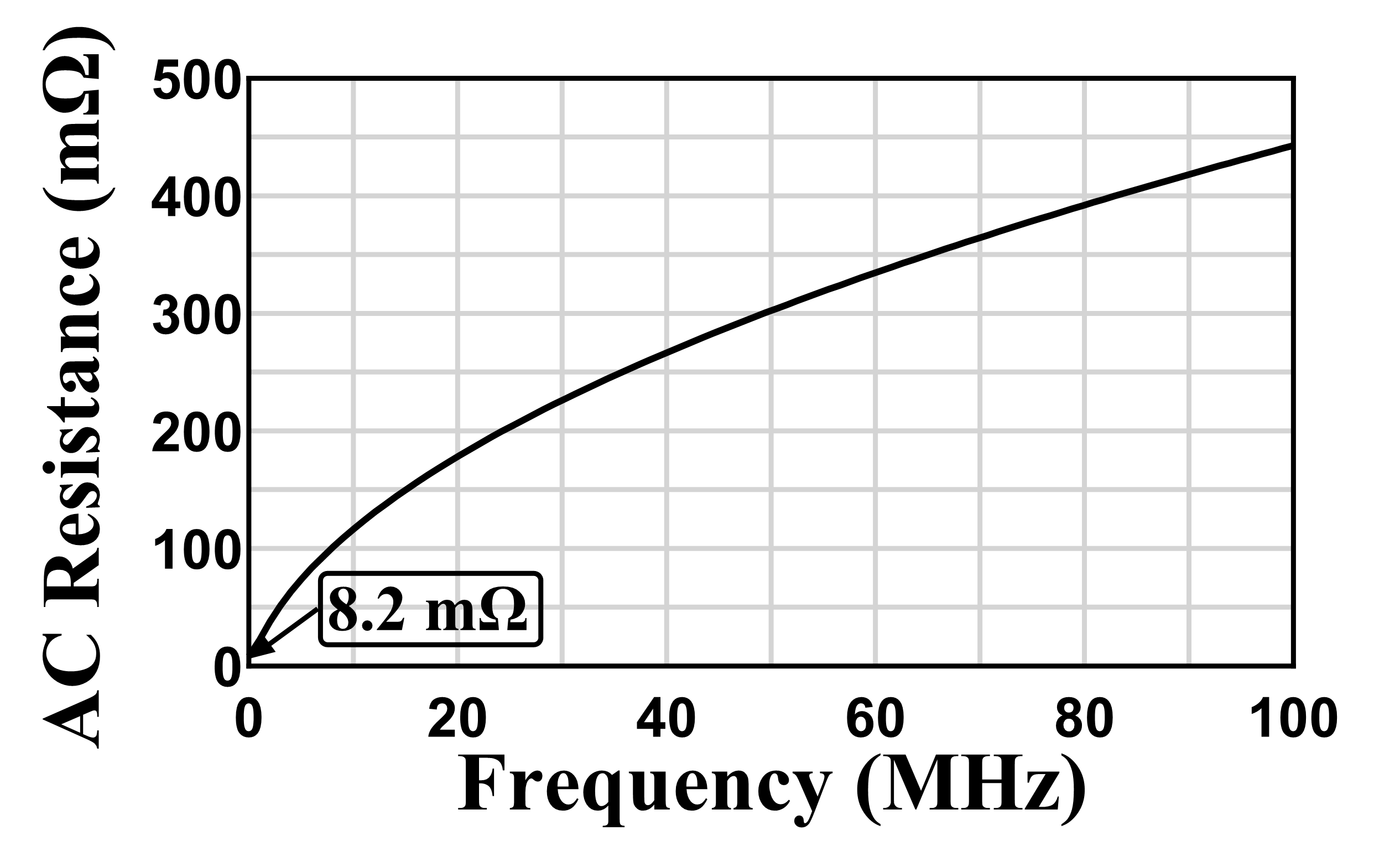}%
\label{ACR_data}}
\vfil\vspace{-10pt}
\subfloat[]{\includegraphics[width=0.24\textwidth]{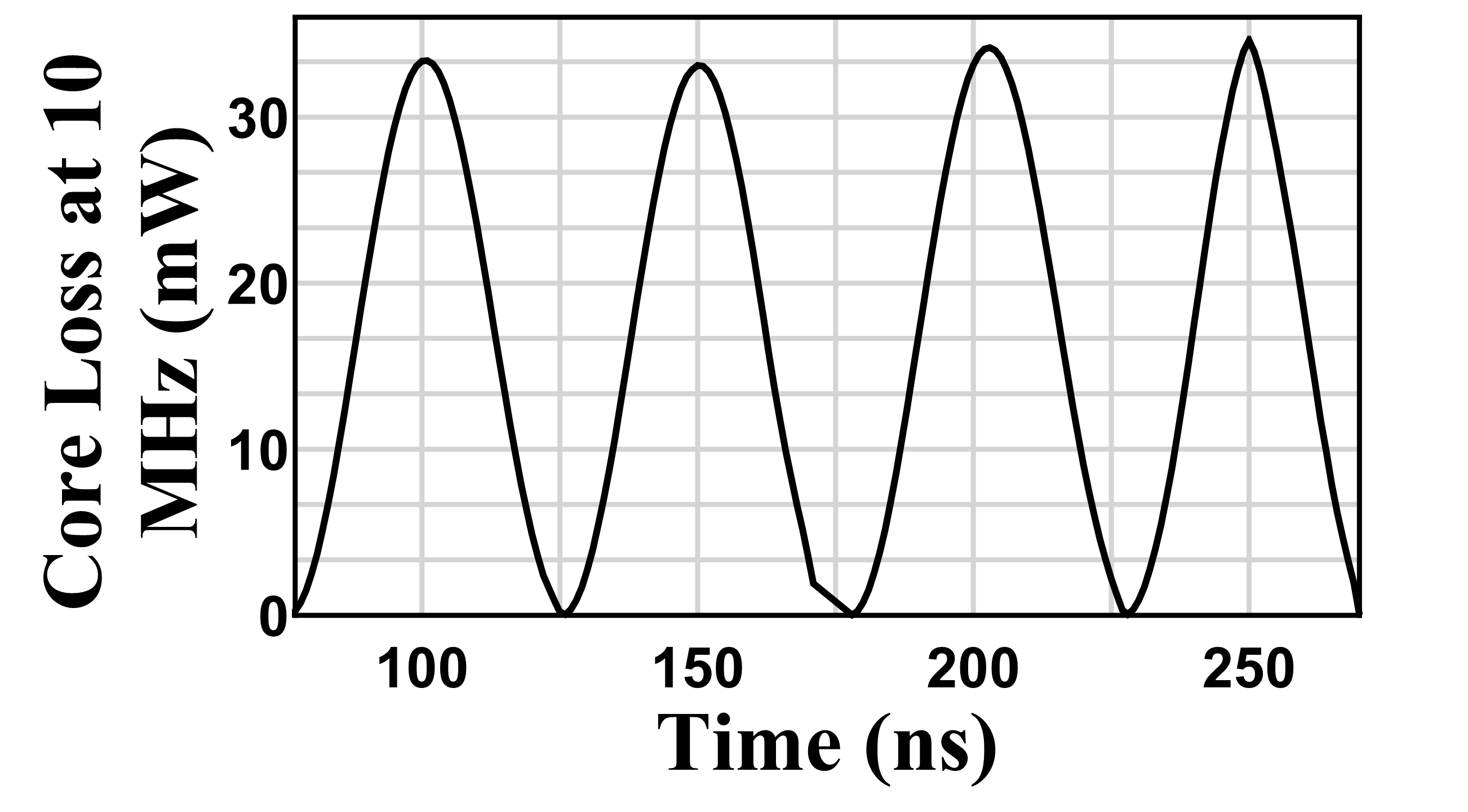}
\label{array_core_loss}}
\hfil
\subfloat[]{\includegraphics[width=0.24\textwidth]{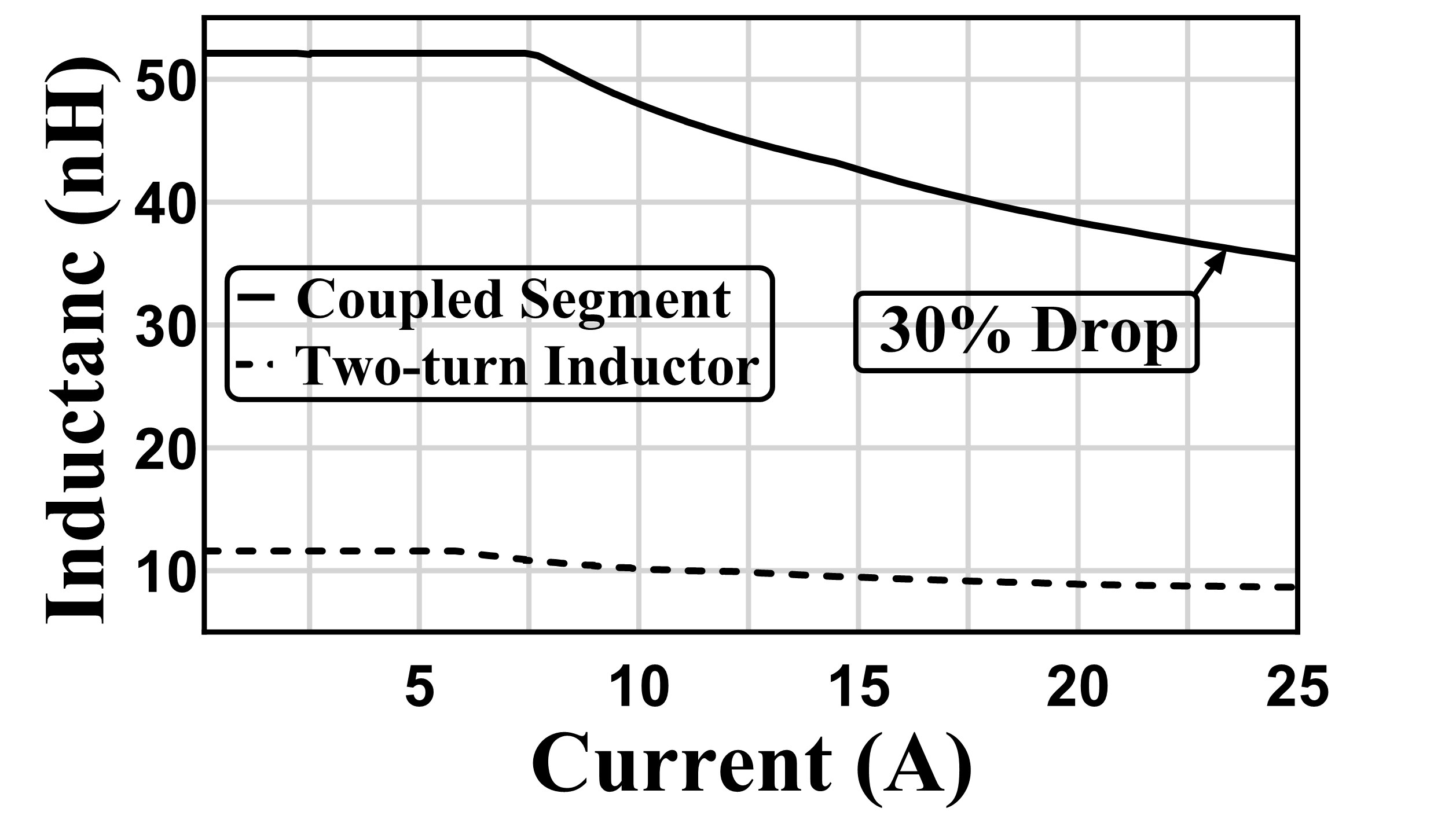}%
\label{Isat}}
\caption{Performance characteristics of a two-turn segment within the proposed coupled array and an optimized two-turn standalone inductor: 
(a) average inductance vs. frequency, 
(b) parasitic resistance (same for both configurations) vs. frequency, (c) core loss of a 10-inductor array at 6 V input and 10 MHz frequency, and
(d) DC inductance vs. DC current.
}
\label{fig:inductor_characteristics}
\end{figure}
}





\begin{table}[]
\centering
\caption{Thermal Analysis Assumptions.}
\label{tab:thermal_table} 
\resizebox{\columnwidth}{!}{
\begin{tabular}{c|l}
\textbf{Parameter} & \textbf{Value} \\
\hline
Inductors per array & 10 \\
Inductor DC current & 2~A \\
Switching frequency & 10~MHz \\
Input voltage & 6~V \\
Free air convection coefficient & 5~W/(m$^{2}\cdot$K)~\cite{ECTC_Gh}\\
Cold plate convection coefficient & 10,000~W/(m$^{2}\cdot$K)~\cite{ghani2024microchannel}\\
Substrate thermal conductivity & 0.76~W/(m$\cdot$K)~\cite{Rogers_TMM10i}\\
Substrate thickness & 1 mm~\cite{avula2024design}\\
\hline
\end{tabular}}
\vspace{-10pt}
\end{table}

\begin{table}[b!]
  \centering
    \caption{Comparison with State-of-the-Art Inductors Operating Within the Range of Interest.}
    \label{tab:inductor_designs}
    \renewcommand{\arraystretch}{1.1} 
    \setlength{\tabcolsep}{2pt}        

    \begin{tabular}{l|l|c|c|c|c|c|c}
      \textbf{Reference} & \textbf{Design} & \textbf{\(\mu_\text{r}\)} & \textbf{\(f_\text{sw}\)} & \textbf{DCR} & \textbf{\(I_\text{sat}\)} & \textbf{L} & \textbf{L\(_\square\)} \\
      & & core & (MHz) & (m\(\Omega\)) & (A) & (nH) & $\big(\frac{\text{nH}}{\text{mm}^2}\big)$ \\
      \hline
      TCPMT'21~\cite{barros2021embedded} & Toroid & 180 & 1 & 23 & 0.1 & 480 & 96 \\
      ECTC'22~\cite{9816521} & Toroid & 30 & 1 & 89 & 2.5 & 420 & 48 \\
      TCPMT'25~\cite{avula2024design} & Single spiral & 90 & 2 & 30 & 5 & 115 & 13 \\
      TCPMT'25~\cite{avula2024design} & Two parallel spiral & 90 & 2 & 85 & 3.5 & 290 & 32 \\
      TCPMT'25~\cite{avula2024design} & Two parallel spiral & 180 & 2 & 85 & 3.5 & 330 & 37 \\
      APEC'14~\cite{burton2014fivr} & FIVR & 1 & 140 & 7 & 8 & 1.2 & 0.6 \\
      ECTC'20~\cite{sankarasubramanian2020magnetic} & Stripline array & N/A & 100 & 12 & 4 & 3 & 6 \\
      ECTC'21~\cite{bharath2021integrated} & Coax & 8.5 & 90 & 12 & 8 & 2.5 & 6.3 \\
      ECTC'25~\cite{ramiectc} & Progressive spiral & 4.7 & 100 & 11 & 24 & 78 & 11.1 \\
      ECTC'25\textsuperscript{*}~\cite{ramiectc}& Progressive spiral & 15 & 100 & 11 &$>$70  & 107 &15.2  \\
      \textbf{This work} & Ten-inductor array & 15 & 10 & 8.2 & 23 & 50 & 250 \\
      \hline
    \end{tabular}
    \begin{tablenotes}
        \item \hspace{-5pt}\textsuperscript{*}Re-simulated with TY-M5 material for fair comparison.
    \end{tablenotes}
\end{table}

\begin{figure}[t!]
\centerline{\includegraphics[scale = 0.4]{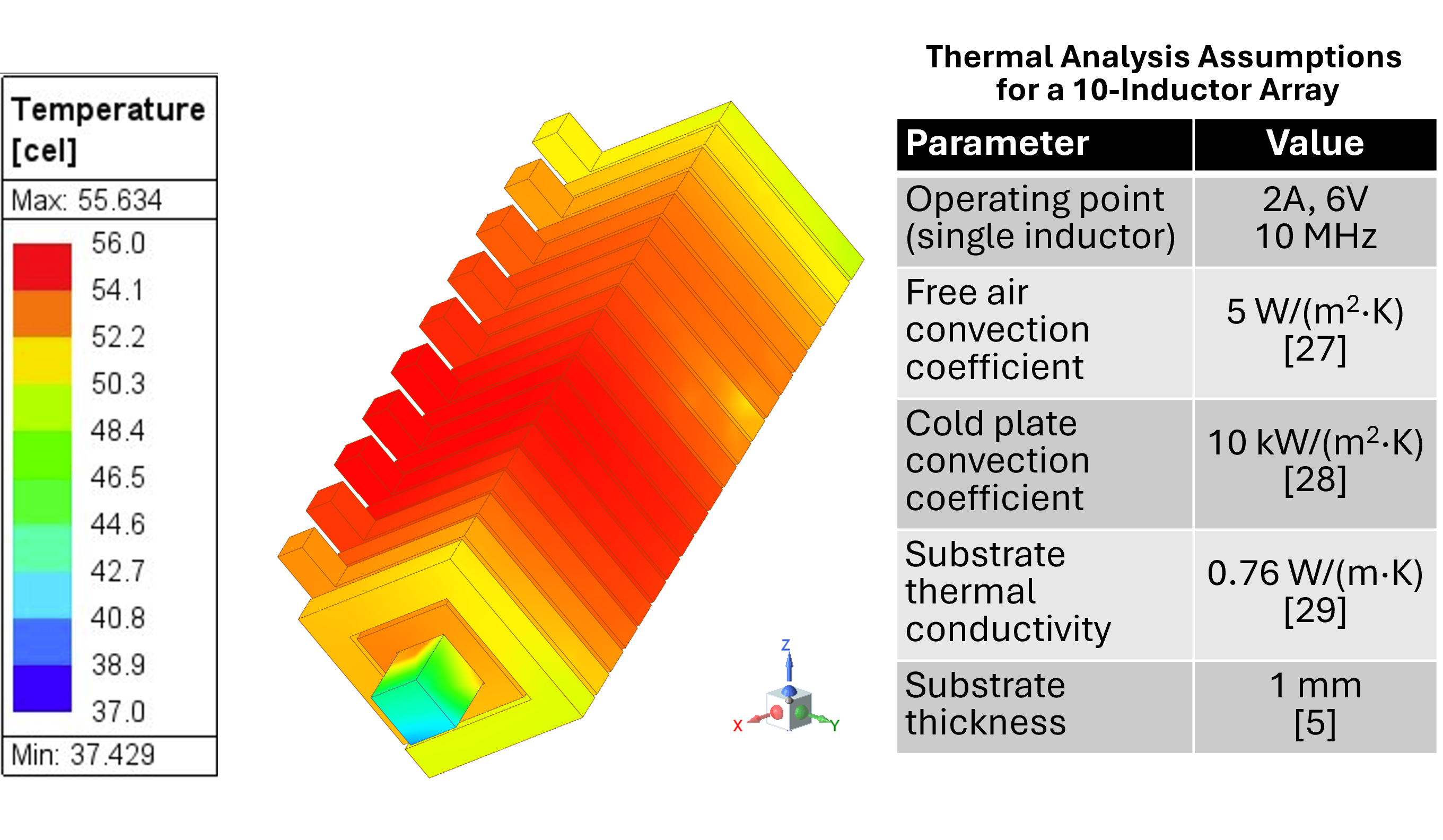}}
\caption{Temperature map of the inductor array at 2 A DC current and 10 MHz frequency.}
\label{thermal_run}
\end{figure}

\subsection{Thermal Analysis}
\label{thermal_analysis_section}
Thermal performance represents a critical challenge for embedded power delivery in HPC systems~\cite{liu2025thermal, rasheedi202528}. Reported cooling approaches include micro-channel cooling~\cite{ding2020novel}, thermal through-silicon vias~\cite{ren2020thermal}, and cold plate attachment~\cite{ghani2024microchannel}. In this work, thermal extraction is modeled in ANSYS Mechanical with a cold plate attached to the bottom surface of the package substrate embedding the inductor array. The assumptions applied in the thermal simulation 
and the resulting temperature distribution are shown in Fig.~\ref{thermal_run}, with a maximum temperature of 55.6$^\circ$C. In this study, the current per inductor is limited to 2~A to achieve this tightly controlled thermal profile; however, with modern cooling technologies, substantially higher current values can be supported \cite{choi2024thermal, choi2025substrate}.
\begin{figure*}[]
    \centering
    \includegraphics[scale=0.6]{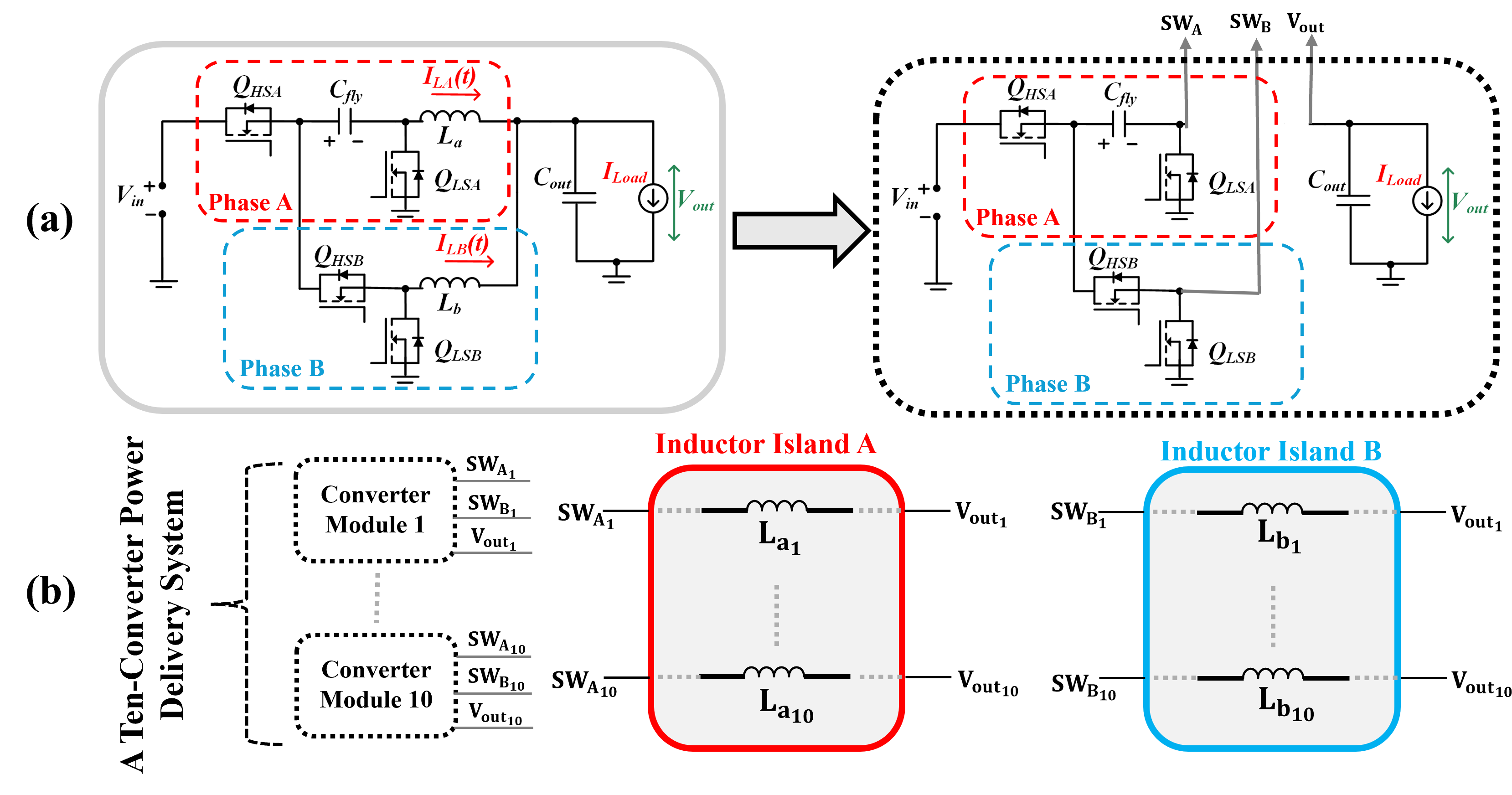}
    \caption{Inductor island-driven power conversion, (a) hybrid converter architecture~\cite{DSCB} and an inductor-less converter module, and 
    (b) two-island circuit block diagram with each island enabled by the proposed ten-inductor array.}
    \vspace{-3pt}
    \label{circuit_block_diagram}
\end{figure*}

\begin{figure}[]
    \centering
    \includegraphics[scale=0.43]{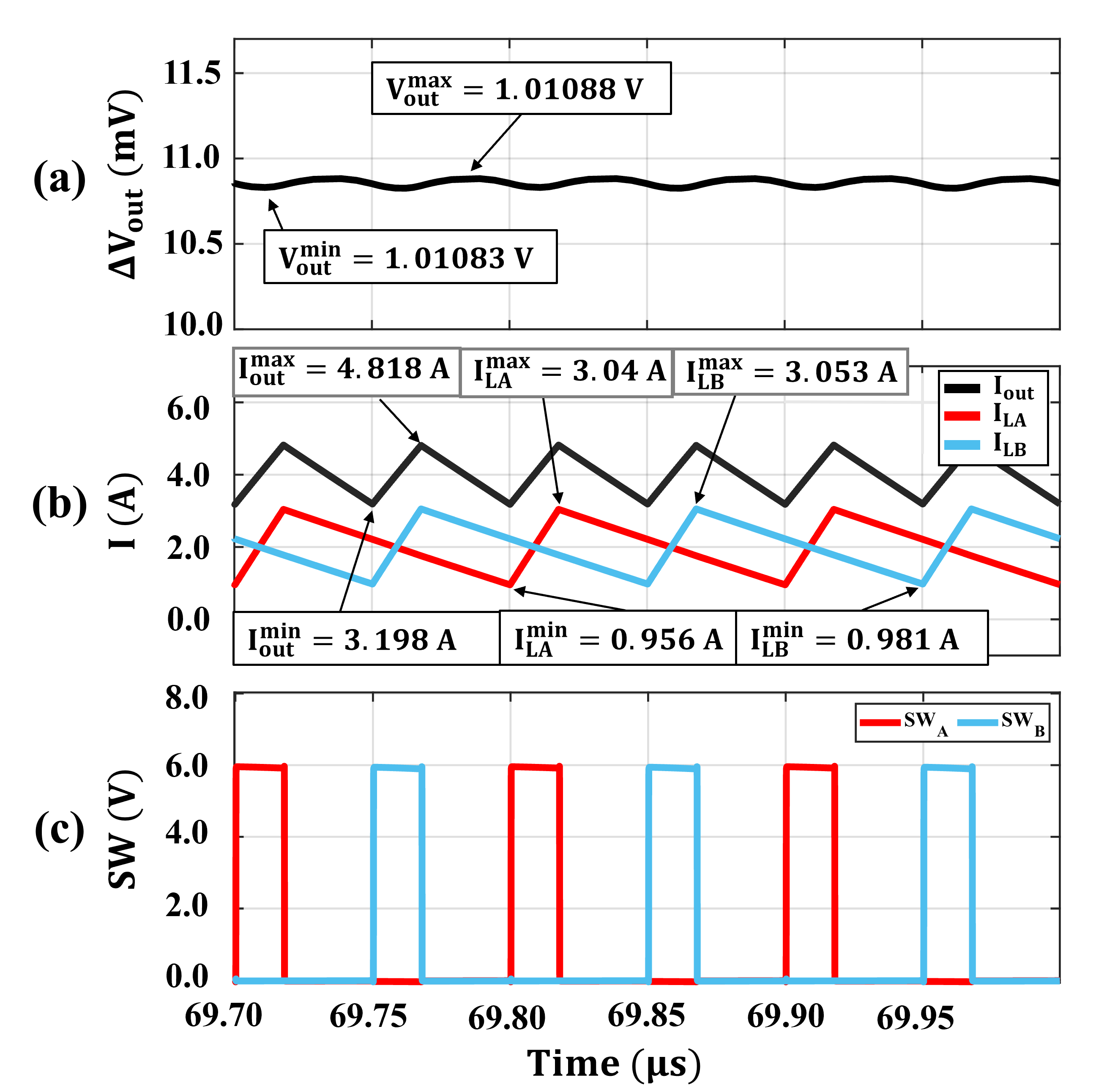}
    \caption{Performance of a single hybrid converter, (a) output voltage, (b) inductor currents ($\text{I}_\text{LA}$ and $\text{I}_\text{LB}$) and output current ($\text{I}_\text{out}$), illustrating steady-state operation and corresponding current ripple characteristics, and (c) switching signals $\text{SW}_\text{A}$ and $\text{SW}_\text{B}$.
    \vspace{-10pt}}
    \label{output_waveforms}
\end{figure}

\begin{figure*}[]
\centering
\subfloat[]{\includegraphics[width=0.5\textwidth]{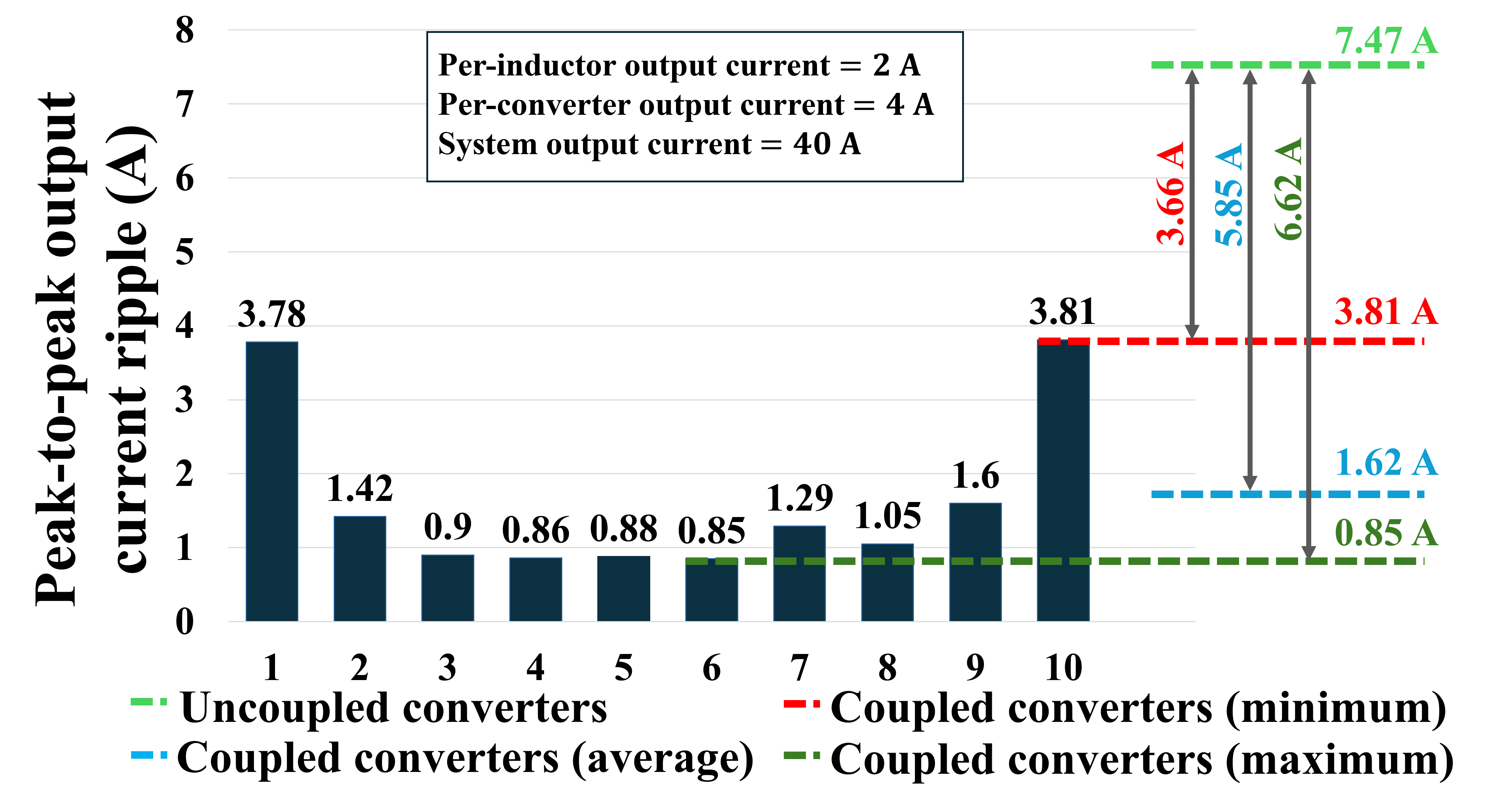}%
\label{p2p_output_ripples_comparison}}
\hfil
\subfloat[]{\includegraphics[width=0.5\textwidth]{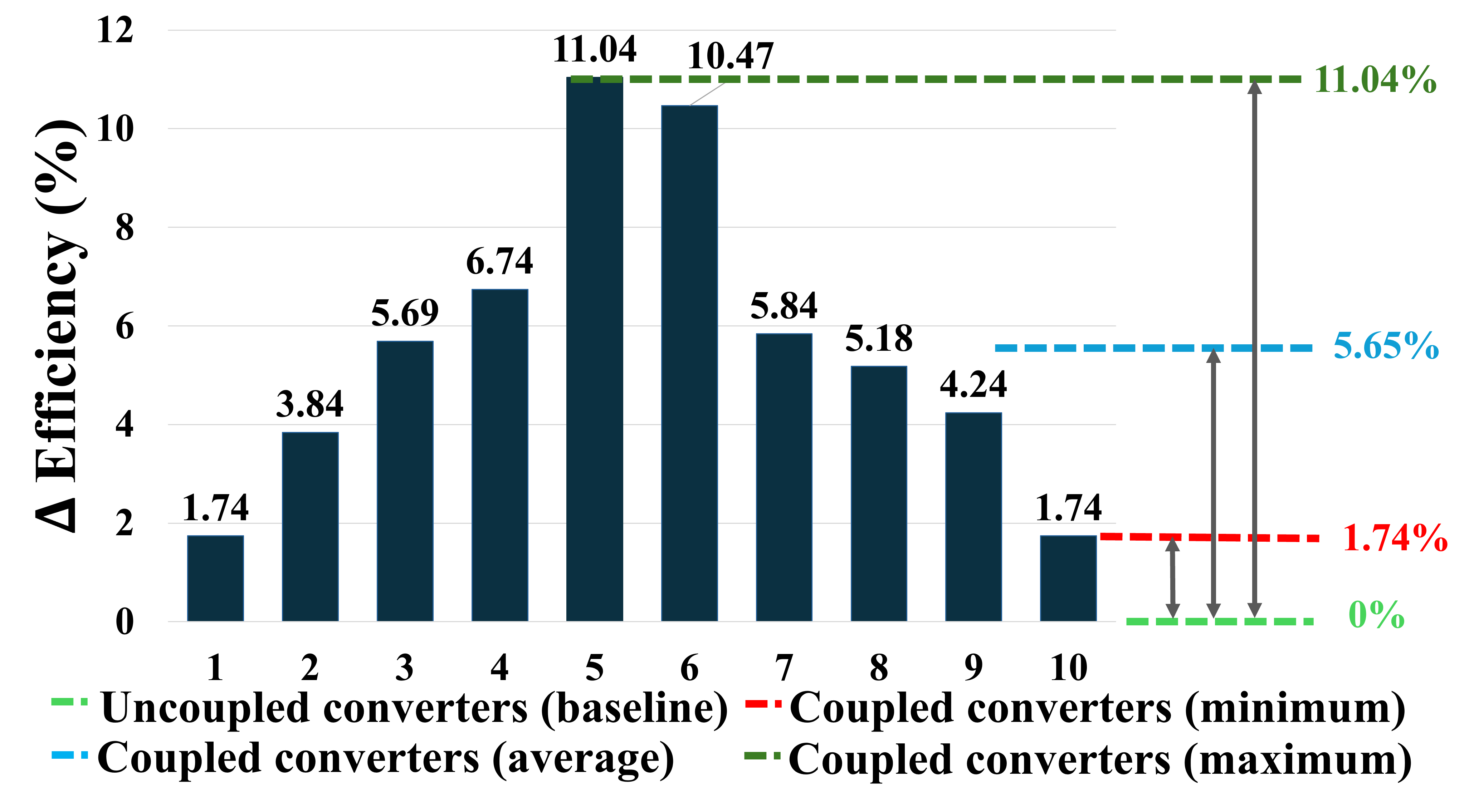}%
\label{efficiencies}}
\caption{\irv{Performance of the individual converters in the proposed coupled power delivery system with ten distributed converters, (a) peak-to-peak output current ripple, and (b) \imv{power efficiency gain. Average, minimum, and maximum performance values across the ten converters are reported on the right as referenced to the baseline system}}}
\vspace{-10pt}
\label{fig:metrics}
\end{figure*}

\subsection{Comparison with State-of-the-Art}
Performance of a single inductor within the proposed inductor array and state-of-the-art inductors operating within the range of interest is presented in Table~\ref{tab:inductor_designs}. Although the inductance of the coupled inductor array is moderate, an excellent inductance density is obtained due to the small footprint of the individual inductors within the array. A high saturation current is achieved, which is well aligned with distributed power delivery in HPC systems, where the total load current of thousands of amperes is supplied by hundreds of parallel VRs. 
Low DCR is another critical performance metric, by which the proposed inductor array is considered favorable compared to state-of-the-art implementations. 

\section{Converter Simulation Results}
\label{Section_4}
To evaluate the circuit-level performance and efficiency of the proposed island-based inductor array architecture, a SPICE model of a ten-inductor array is extracted from ANSYS Maxwell 3D. 

\subsection{System-Level Converter Simulations}

A system of ten parallel-connected hybrid dual-phase switched-capacitor buck converters is designed in Cadence Virtuoso, based on~\cite{DSCB} and the coupled inductor arrays. \imv{Both the coupled and uncoupled systems are evaluated using the same converter topology, operating conditions, and loss models, such that the observed differences are solely attributable to the inductor implementation.}

A single hybrid converter is shown in Fig.~\ref{circuit_block_diagram}a. 
The converter steps a 12\,V input down to 1\,V through two stages: 1) the SC stage halves the input voltage with a single flying capacitor, reducing it from 12\,V to 6\,V at the switching nodes ($\text{SW}_\text{A}$ and $\text{SW}_\text{B}$), 2) the dual-phase buck stage further steps the voltage down from 6\,V to 1\,V with a duty cycle of 1/6. The target output current is 4\,A per converter, limited to 2\,A per inductor due to thermal constraints.

The overall circuit-level architecture and interconnections of the island-based power delivery system are presented in Fig.~\ref{circuit_block_diagram}b. Two inductor islands, Island A and Island B, are operated in phases~A and~B, respectively. Each of the ten converter modules is connected to both islands, utilizing the ten-inductor array \(L_{A}\) from Island~A and the ten-inductor array \(L_{B}\) from Island~B.
The design parameters of the system are summarized in Table~\ref{tab:system_params}, and the selected components are provided in Table~\ref{tab:component_selection}.

\begin{table}
\caption{System Parameters of the Overall Distributed Power Conversion System.}
\label{tab:system_params}
\centering
\resizebox{\columnwidth}{!}{
\begin{tabular}{l|l}
\centering \textbf{{Parameter}} & \textbf{{Value}} \\
\hline
System input/output voltage  & $\text{\textit{V}}_\text{in}$ = 12 V; $\text{\textit{V}}_\text{out}$ = 1 V \\
System switching frequency & $f_{\text{sw}}$ = 10 MHz \\
System/converter output current & $\text{\textit{I}}_\text{sys}$ = 40 A; $\text{\textit{I}}_\text{out}$ = 4 A \\
Inductor output current & $\text{\textit{I}}_\text{LA}$ = $\text{\textit{I}}_\text{LB}$ = 2 A \\
\hline
\end{tabular}}
\end{table}

\begin{table}
\caption{Selected Components for the Individual Hybrid Power Converters.}
\label{tab:component_selection}
\centering
\resizebox{\columnwidth}{!}{
\begin{tabular}{l|l}
\centering \textbf{Component Type} & \textbf{Selected Component} \\
\hline
\centering High-side switch & EPC2040 \\
\centering Low-side switch & EPC2216 \\
\centering Flying capacitor \hspace{15pt} & $C_{\text{fly}} = 3.3$~\textmu F, ESR =  3\,m$\Omega$ \hspace{15pt} \\
\centering Output capacitance & $C_\text{out} = 100$~\textmu F \\
\hline
\end{tabular}}
\end{table}
%


To establish a baseline for performance evaluation of the proposed power delivery system, a single two-phase hybrid converter is co-designed with two optimized, two-turn uncoupled inductors (see the progressively widened spiral inductor in Table~\ref{tab:inductor_designs}). 
Consequently, the overall \texttt{baseline} power delivery system is constructed from ten such converters connected in parallel. 
In contrast, the \texttt{proposed} coupled power delivery system is implemented with ten parallel converter modules, each employing two inductor segments drawn from the two coupled inductor arrays.

\imv{In vertically integrated power delivery systems, the achievable current density is fundamentally constrained by the component with the largest footprint, which in this work corresponds to the power switches. Accordingly, the system-level current density is estimated to be 1 A/mm$^2$ based on the switch-dominated area. In contrast, the proposed inductor array supports a significantly higher intrinsic current density of approximately 10 A/mm$^2$, indicating that the inductors are not the limiting factor. Therefore, the proposed architecture is best interpreted as a scalable, density-driven power delivery solution, where the total deliverable current can be increased through spatial replication of unit cells while maintaining constant current density.}

Both two-phase systems with ten parallel-connected converters are simulated in Cadence Virtuoso at $10\,\text{MHz}$. The 40-A baseline system achieves an efficiency of 80.46\% for each of the ten uncoupled parallel converters, with a peak-to-peak output current ripple of 7.47~A per converter at a total output current of 4~A. The steady-state average performance across the ten converters of the proposed coupled system is shown in Fig.~\ref{output_waveforms}. Intuitively, 
converters utilizing inductors near the center (e.g., Modules 5, 6 in Fig.~\ref{circuit_block_diagram}) of the array benefit more from positive magnetic coupling, resulting in improved performance compared to the edge converters (e.g., Modules 1, 10 in Fig.~\ref{circuit_block_diagram}). Note that the number of top-performing modules increases with an increasing number of coupled inductors. Thus, the size of the coupled inductor array should be determined based on the specific profile of the HPC system (e.g., number of high-performance chiplets and lower-power chiplets).

The distribution of the output peak-to-peak current ripple across all ten coupled converters is presented in Fig.~\ref{p2p_output_ripples_comparison}. For reference, the ripple of the uncoupled baseline system is included for direct comparison. The proposed coupled architecture achieves a reduction in output current ripple of at least 3.66\,A (9.15\% of the system output current), reaching up to 6.62\,A (16.55\%), with an average reduction of 5.85\,A (14.625\%).
In the remainder of the paper, average performance characteristics (see Fig.~\ref{output_waveforms}) are used for comparison with other inductors.

The simulated converter waveforms in Fig.~\ref{output_waveforms} confirm steady state operation at approximately 1~V output. 
The two inductor currents share the 4~A converter load and exhibit of approximately 2.08\,A peak-to-peak ripple each while phase interleaving reduces output current ripple to approximately 1.62\,A, corresponding to 40\% of the nominal converter output current. The resulting voltage ripple is only 50 \textmu V, indicating that the selected output capacitance effectively suppresses ripple at the converter output and falls within the typical design range for high-current voltage regulators~\cite{current_ripple_minimization}.

\irv{\subsection{Efficiency and Loss Analysis}}
%

\irv{To more accurately evaluate the efficiency, first-order loss mechanisms not fully captured in simulations under ideal gate excitation and simplified parasitic assumptions are explicitly included: (i) GaN dead-time conduction loss and (ii) gate-drive power loss. Based on the I–V characteristics reported for the EPC2040 (HS) and EPC2216 (LS) devices \cite{EPC2040_datasheet, EPC2216_datasheet}, the estimated dead-time loss is 0.168 W for a dead time of 1 ns \cite{DT_1, DT_2}. The gate-drive loss, obtained from an analytical gate-drive model using datasheet parameters and validated with the EPC loss estimator, is 0.167 W at a switching frequency of 10 MHz \cite{ECTC-salma, EPC2040_datasheet, EPC2216_datasheet, EPC_GaNPowerBench_2025}.

Together, these mechanisms introduce an additional high-frequency loss of 0.335~W beyond the simulated losses at the operating point $\text{P}_{\text{out}} = 4~\text{W}$ ($\text{V}_{\text{o}} = 1~\text{V}$, $\text{I}_{\text{o}} = 4~\text{A}$). Additional routing $\text{I}^2\text{R}$ losses are conservatively estimated to be 20–30 mW and therefore do not materially impact the overall efficiency \cite{krishnakumar2024_GLSVLSI, PG_HPC}. This estimate follows from the compact converter layout enabled by vertical stacking of components, which limits horizontal current paths to approximately 1–2~mm within the $1~\text{mm}^2$ VR footprint. With standard 1-oz copper routing and parallel current-sharing traces, the resulting dissipation remains much smaller than the dominant dead-time and gate-drive losses.

Under soft-switching conditions enabled by the buck inductors, a substantial portion of the $\text{C}_\text{oss}$ commutation energy is recovered. Even assuming non-ideal soft switching (e.g., reduction factor of 0.9), the residual $\text{C}_\text{oss}$ loss is approximately one order of magnitude smaller than the dead-time and gate-drive losses \cite{ECTC-salma, EPC2040_datasheet, EPC2216_datasheet}. Accordingly, both routing and residual $\text{C}_\text{oss}$ losses are treated as secondary contributions and are not included in the final adjusted efficiency calculation.

\imv{The power efficiency gain of the proposed power delivery system with ten distributed converters compared to the baseline system is shown in Fig. \ref{efficiencies}}. While the absolute efficiency depends on implementation-specific parasitics and packaging configuration, the comparative efficiency improvement between the coupled and uncoupled architectures remains consistent under practical loss assumptions. Since the same dead-time, gate-drive, and routing mechanisms apply to both cases, the relative performance benefit of the proposed shared-core inductor islands is preserved under realistic loss modeling.} \imv{The efficiency results are intended to provide a relative comparison between the proposed and baseline architectures under identical modeling assumptions, rather than to claim absolute efficiency values for a specific implementation.}

\imv{For the lowest-performing coupled converters (Modules 1 and 10), the efficiency improves by 1.74\% relative to the uncoupled baseline, representing the minimum observed benefit of enabling magnetic coupling. In contrast, the best-performing converters (Modules 5 and 6) achieve an efficiency improvement of 11.04\% over the uncoupled case. Averaged across all ten converters, the proposed coupled system provides a 5.65\% efficiency gain compared with the baseline system.}
These results confirm that the efficiency improvement is directly correlated with the magnetic-coupling strength within the shared-core inductor island.

\vspace{10pt}
\section{Conclusion}
\label{Conclusion}
A novel package-embedded inductor architecture and an inductance-island based power delivery methodology are proposed for high-efficiency vertical power delivery in high-performance computing systems. The topology employs an array of tightly coupled, shared-core inductors organized into phase-specific islands to simultaneously enhance inductance density and current-handling capability while maintaining a compact footprint. The individual two-turn vertical segments of the coupled inductor arrays are optimized for maximum quality factor, achieving inductance density of up to 250~nH/mm$^2$ and a current density of 10~A/mm$^2$, with an inductor efficiency of 97.4\% at 2~A and 50~MHz. Compared to uncoupled configuration, coupled inductor arrays provide a significant efficiency improvement in a distributed 10-module converter designed in Virtuoso, achieving an average gain of \irv{5.65\%} and up to \irv{11.04\%} higher overall conversion efficiency at 40~A. These results demonstrate the effectiveness of same-phase coupled inductor arrays for multi-phase power delivery. 
The proposed approach is evaluated using physics-based electromagnetic modeling, thermal analysis, and circuit-level co-design, providing a comparative assessment of the architecture under realistic operating assumptions.

Electrical and thermal analyses further validate the design under realistic operating conditions.
\imv{ 
The proposed methodology is combines physics-based electromagnetic modeling with circuit-level co-simulation to evaluate coupled inductor arrays and distributed VPD architectures under practical operating constraints. The presented results provide a comparative architectural assessment of the proposed approach, while
experimental realization of fully integrated package-embedded implementations is deferred to future work.}
The proposed architecture provides a scalable and energy-efficient solution for high-performance voltage regulators in advanced heterogeneous packaging.


\section*{Acknowledgment}

This work was supported by the Center for Heterogeneous Integration of Micro Electronic Systems (CHIMES), one of seven centers in Joint University Microelectronics Program (JUMP) 2.0, a Semiconductor Research Corporation (SRC) program sponsored by the Defense Advance Research Project Agency (DARPA).

\bibliographystyle{IEEEtran}
\bibliography{Refs}

@INPROCEEDINGS{1,
  author={Krishnakumar, Sriharini and Partin-Vaisband, Inna},
  booktitle={IEEE International System-on-Chip Conference (SOCC)}, 
  title={Vertical Power Delivery for Emerging Packaging and Integration Platforms---Power Conversion and Distribution}, 
  year={2023},
  volume={},
  number={},
  pages={1-6},
  doi={10.1109/SOCC58585.2023.10256973}}

@inproceedings{2,
  title={Designing of Multilayer Planar Spiral Air-Core Inductor for Power Electronic Applications},
  author={Khakroei, Mohammad and Mostafaei, Mohsen and Arefian, Mansour and Rezaei-Zare, Afshin and Zarmehri, Majid Najafi},
  booktitle={International Conference on Electrical Engineering (ICEE)},
  pages={71--77},
  year={2022},
  organization={IEEE}
}

@ARTICLE{3,
  author={Nishad, Praween Kumar and Ghosh, Debapratim},
  journal={IEEE Transactions on Magnetics}, 
  title={Two-Layer Planar Rectangular Inductors and Their Electrical Models for Increased Inductance Density and Figure-of-Merit}, 
  year={2023},
  volume={59},
  number={3},
  pages={1-8},
  doi={10.1109/TMAG.2023.3239666}}

@article{grover1946inductance,
  title={Inductance Calculations--New York},
  author={Grover, FW},
  journal={Van Norstrand Company inc},
  year={1946}
}

@inproceedings{rasheedi2024embedded,
  title={An Embedded Multi-Layer Spiral Square Inductor for Integrated Power Delivery-Physical Design and Analytical Models},
  author={Rasheedi, Rami and Partin-Vaisband, Inna},
  booktitle={Proceedings of the Great Lakes Symposium on VLSI},
  pages={370--375},
  year={2024}
}

@article{radhakrishnan2021power,
  title={Power delivery for high-performance microprocessors—challenges, solutions, and future trends},
  author={Radhakrishnan, Kaladhar and Swaminathan, Madhavan and Bhattacharyya, Bidyut K},
  journal={IEEE Transactions on Components, Packaging and Manufacturing Technology},
  volume={11},
  number={4},
  pages={655--671},
  year={2021},
  publisher={IEEE}
}

@article{he2023soft,
  title={Soft magnetic materials for power inductors: State of art and future development},
  author={He, Jiayi and Yuan, Han and Nie, Min and Guo, Hai and Yu, Hongya and Liu, Zhongwu and Sun, Rong},
  journal={Materials Today Electronics},
  pages={100066},
  year={2023},
  publisher={Elsevier}
}

@article{avula2024design,
  title={Design and demonstration of dual-core spiral package-embedded inductors for Integrated Voltage Regulators},
  author={Avula, Venkatesh and Murali, Prahalad and Swaminathan, Madhavan},
  journal={IEEE Transactions on Components, Packaging and Manufacturing Technology},
  year={2025},
  publisher={IEEE}
}

@INPROCEEDINGS{9816521,
  author={Murali, Prahalad and Avula, Venkatesh and Ahmed, Marisa and Losego, Mark D. and Swaminathan, Madhavan and Alvarez, Claudio and Oishi, Yusuke and Uemura, Tomohito and Nagatsuka, Ryo and Watanabe, Naoki},
  booktitle={IEEE Electronic Components and Technology Conference (ECTC)}, 
  title={Fabrication and Characterization of Package Embedded Inductors for Integrated Voltage Regulators}, 
  year={2022},
  volume={},
  number={},
  pages={301-305},
  doi={10.1109/ECTC51906.2022.00056}}

@inproceedings{burton2014fivr,
  title={{FIVR}—Fully integrated voltage regulators on 4th generation {Intel}{\textregistered} {Core™ SoCs}},
  author={Burton, Edward A and Schrom, Gerhard and Paillet, Fabrice and Douglas, Jonathan and Lambert, William J and Radhakrishnan, Kaladhar and Hill, Michael J},
  booktitle={IEEE Applied Power Electronics Conference and Exposition (APEC)},
  pages={432--439},
  year={2014}
}

@inproceedings{sankarasubramanian2020magnetic,
  title={Magnetic inductor arrays for intel{\textregistered} fully integrated voltage regulator ({FIVR}) on 10\textsuperscript{th} generation {Intel{\textregistered} Core™ SoCs}},
  author={Sankarasubramanian, Malavarayan and Radhakrishnan, Kaladhar and Min, Yongki and Lambert, William and Hill, Michael J and Dani, Ashay and Mesch, Ryan and Wojewoda, Leigh and Chavarria, Jose and Augustine, Anne},
  booktitle={IEEE Electronic Components and Technology Conference (ECTC)},
  pages={399--404},
  year={2020},
}

@inproceedings{bharath2021integrated,
  title={Integrated voltage regulator efficiency improvement using coaxial magnetic composite core inductors},
  author={Bharath, Krishna and Radhakrishnan, Kaladhar and Hill, Michael J and Chatterjee, Prithwish and Hariri, Haifa and Venkataraman, Srikrishnan and Do, Huong T and Wojewoda, Leigh and Srinivasan, Sriram},
  booktitle={IEEE Electronic Components and Technology Conference (ECTC)},
  pages={1286--1292},
  year={2021},
}

@article{barros2021embedded,
  title={Embedded inductors using composite magnetic materials for 12--1-{V} integrated voltage regulators},
  author={Barros, Claudio Alvarez and Murali, Prahalad and Swaminathan, Madhavan and Yusuke, Oishi and Junichi, Takashiro and Ryo, Nagatsuka and Watanabe, Naoki},
  journal={IEEE Transactions on Components, Packaging and Manufacturing Technology},
  volume={11},
  number={12},
  pages={2183--2192},
  year={2021},
  publisher={IEEE}
}

@INPROCEEDINGS{ramiectc,
  author={Rasheedi{}, Rami{} and Partin-Vaisband, Inna},
  booktitle={IEEE Electronic Components and Technology Conference (ECTC)}, 
  title={High Aspect Ratio Spiral Inductor with Progressive Turn Widths for Embedded Power Converters}, 
  year={2025},
  volume={},
  number={},
  pages={2271-2277},
  doi={10.1109/ECTC51687.2025.00386}}

@INPROCEEDINGS{apec2025,
  author={Kandeel, Youssef and Ye, Liang and Flannery, John and Mathúna, Cian O and Sai, Ranajit and O’Driscoll, Séamus and Tsuchida, Takayuki and Terauchi, Naoya and Kishimoto, Sumiaki and Hiraoka, Toshio and Nagano, Masanori},
  booktitle={IEEE Applied Power Electronics Conference and Exposition (APEC)}, 
  title={High-Efficiency {PCB}-Embeddable Inductor for Vertical Power {IVR} Applications}, 
  year={2025},
  volume={},
  number={},
  pages={285-290},
  doi={10.1109/APEC48143.2025.10977275}}

@INPROCEEDINGS{10509290,
  author={Gan, Houle and Jiang, Shuai and Teng, Sue and Yamamoto, Shin and Chivukula, Venkata and Edwards, Bill and Chung, Chee and Chen, Jason and Mohideen, Mushafik and Sizikov, Gregory and Li, Xin},
  booktitle={IEEE Applied Power Electronics Conference and Exposition (APEC)}, 
  title={Vertical Power Delivery for 1000 {Amps} Machine Learning {ASICs}}, 
  year={2024},
  volume={},
  number={},
  pages={906-909},
  doi={10.1109/APEC48139.2024.10509290}}

@article{DSCH,
  title={High-Efficiency Nonisolated Converter With Very High Step-Down Conversion Ratio},
  author={Kirshenboim, Or and Peretz, Mor Mordechai},
  journal={IEEE Transactions on Power Electronics},
  volume={32},
  number={5},
  pages={3683--3690},
  year={2016},
  publisher={IEEE}
}

@inproceedings{DPMIH44,
  title={A Regulated {48V-to-1V/100A} 90.9\%-Efficient Hybrid Converter for {POL} Applications in Data Centers and Telecommunication Systems},
  author={Das, Ratul and Le, Hanh-Phuc},
  booktitle={IEEE Applied Power Electronics Conference and Exposition (APEC)},
  pages={1997--2001},
  month={Mar},
  year={2019}
}

@inproceedings{DSCB,
  title={A {5 MHz}, {12 V}, {10 A}, Monolithically Integrated Two-Phase Series Capacitor Buck Converter},
  author={Shenoy, Pradeep S and Lazaro, Orlando and Ramani, Ramanathan and Amaro, Mike and Wiktor, Wlodek and Khayat, Joseph and Lynch, Brian},
  booktitle={IEEE Applied Power Electronics Conference and Exposition (APEC)},
  pages={66--72},
  year={2016},
  month={Mar}
}

@INPROCEEDINGS{ECTC-salma,
  author={Abdelzaher, Salma and Gharib, Mohamed and Partin-Vaisband, Inna},
  booktitle={IEEE Electronic Components and Technology Conference (ECTC)}, 
  title={Hybrid Voltage Regulators for High Performance Computing: Analytical Models and Design Methodology}, 
  year={2025},
  volume={},
  number={},
  pages={2286-2292},
note={\text{\href{10.1109/ECTC51687.2025.00388}{doi:10.1109/ECTC51687.2025.00388}}},
     }

@inproceedings{DPMIH62,
  title={Multiphase Control for Robust and Complete Soft-Charging Operation of Dual Inductor Hybrid Converter},
  author={Xie, Tianshi and Das, Ratul and Seo, Gab-Su and Maksimovic, Dragan and Le, Hanh-Phuc},
  booktitle={IEEE Applied Power Electronics Conference and Exposition (APEC)},
  pages={1--5},
  year={2019},
  month={Mar}
}

@INPROCEEDINGS{ECTC_Gh,
  author={Gharib, Mohamed and Partin-Vaisband, Inna},
  booktitle={IEEE  Electronic Components and Technology Conference (ECTC)}, 
  title={Efficient Scalable Thermoelectric Modeling of High-Frequency Cylindrical Interconnects for Heterogeneous Package Arrays}, 
  year={2025},
  volume={},
  number={},
  pages={2278-2285},
  doi={10.1109/ECTC51687.2025.00387}}

@ARTICLE{8385217,
  author={Le, Hoa Thanh and Nour, Yasser and Pavlovic, Zoran and O'Mathúna, Cian and Knott, Arnold and Jensen, Flemming and Han, Anpan and Kulkarni, Santosh and Ouyang, Ziwei},
  journal={IEEE Transactions on Power Electronics}, 
  title={High-Q Three-Dimensional Microfabricated Magnetic-Core Toroidal Inductors for Power Supplies in Package}, 
  year={2019},
  volume={34},
  number={1},
  pages={74-85},
  doi={10.1109/TPEL.2018.2847439}}

@INPROCEEDINGS{9124544,
  author={Selvaraj, S. Lawrence and Haug, Martin and Cheng, Chor Shu and Dinulovic, Dragan and Peng, Lulu and Shafey, Khaled El and Ali, Zishan and Shousha, Mahmoud and Ng, Yong Chau and Aziz Yosokumoro, Nur and Lehmann, Lothar and Wieland, Marcel},
  booktitle={IEEE Applied Power Electronics Conference and Exposition (APEC)}, 
  title={On-Chip Thin Film Inductor for High Frequency DC-DC Power Conversion Applications}, 
  year={2020},
  volume={},
  number={},
  pages={176-180},
  doi={10.1109/APEC39645.2020.9124544}}

@INPROCEEDINGS{9097695,
  author={Murphy, Ruaidhri and Pavlovic, Zoran and McCloskey, Paul and O Mathuna, Cian and O'Driscoll, Seamus and Weidinger, Gerald},
  booktitle={International Conference on Integrated Power Electronics Systems}, 
  title={{PCB} Embedded Toroidal Inductor for 2{MHz} Point-of-Load Converter}, 
  year={2020},
  volume={},
  number={},
  pages={1-5},
  doi={}}

@article{ghani2024microchannel,
  title={Microchannel heat sinks—A comprehensive review},
  author={Ghani, Usman and Wazir, Muhammad Anas and Akhtar, Kareem and Wajib, Mohsin and Shaukat, Shahmir},
  journal={Electronic Materials},
  volume={5},
  number={4},
  pages={249--292},
  year={2024},
  publisher={MDPI}
}

@online{Rogers_TMM10i,
  author       = {{Rogers Corporation}},
  title        = {{TMM}\textsuperscript{\textregistered} 10i Laminates},
  year         = {2025},
  url          = {https://www.rogerscorp.com/advanced-electronics-solutions/tmm-laminates/tmm-10i-laminates}
}

@article{hinov2023design,
  title={Design considerations of multi-phase buck dc-dc converter},
  author={Hinov, Nikolay and Grigorova, Tsvetana},
  journal={Applied Sciences},
  volume={13},
  number={19},
  pages={11064},
  year={2023},
  publisher={MDPI}
}

@article{liu2025thermal,
  title={Thermal optimization of dual-sided embedded liquid cooling for high-power-density {3D HPC} architectures},
  author={Liu, Yunting and Fu, Rong and Su, Meiying and Li, Jun and Chen, Chuan and Liu, Fengman},
  journal={Microelectronics Journal},
  volume={161},
  pages={106714},
  year={2025},
  publisher={Elsevier}
}

@article{ding2020novel,
  title={A novel thermal management scheme for {3D-IC} chips with multi-cores and high power density},
  author={Ding, Bin and Zhang, Zhi-Hao and Gong, Liang and Xu, Ming-Hai and Huang, Zhao-Qin},
  journal={Applied thermal engineering},
  volume={168},
  pages={114832},
  year={2020},
  publisher={Elsevier}
}

@article{ren2020thermal,
  title={Thermal {TSV} optimization and hierarchical floorplanning for {3-D} integrated circuits},
  author={Ren, Zongqing and Alqahtani, Ayed and Bagherzadeh, Nader and Lee, Jaeho},
  journal={IEEE Transactions on Components, Packaging and Manufacturing Technology},
  volume={10},
  number={4},
  pages={599--610},
  year={2020},
  publisher={IEEE}
}

@ARTICLE{current_ripple_minimization,
  author={Dutta, Soham and Johnson, Brian},
  journal={IEEE Transactions on Power Electronics}, 
  title={A Practical Digital Implementation of Completely Decentralized Ripple Minimization in Parallel-Connected DC–DC Converters}, 
  year={2022},
  volume={37},
  number={12},
  pages={14422-14433},
  doi={10.1109/TPEL.2022.3197674}}

@misc{ieeeHIR2023power,
  author       = {{IEEE Electronics Packaging Society (EPS)}},
  title        = {Heterogeneous Integration Roadmap: Chapter 10 -- Integrated Power Electronics ({HIR} 2023 Edition)},
  howpublished = {[Online] Available: \url{https://eps.ieee.org/images/files/HIR_2023/ch10_power.pdf}},
  year         = {2023}
}

@techreport{src-mapt-roadmap-2023,
  author       = {{Semiconductor Research Corporation (SRC)}},
  title        = {{MAPT} Microelectronics and Advanced Packaging Technologies Roadmap (Version 4)},
  year         = {2023},
  month        = mar,
  url          = {https://srcmapt.org/wp-content/uploads/2024/03/SRC-MAPT-Roadmap-2023-v4.pdf}
}

@inproceedings{choi2024thermal,
  title={Thermal Analysis of High Current Vertical Power Delivery Network with Embedded Microchannel Cooling},
  author={Choi, Mingeun and Krishnakumar, Sriharini and Khorasani, Ramin Rahimzadeh and Partin-Vaisband, Inna and Sharma, Rohit and Swaminathan, Madhavan and Kumar, Satish},
  booktitle={IEEE Intersociety Conference on Thermal and Thermomechanical Phenomena in Electronic Systems (ITherm)},
  pages={1--8},
  year={2024}
}

@article{choi2025substrate,
  title={Substrate-Embedded Microfluidic Cooling of Distributed Vertical Power Delivery Architectures for High-Performance Computing Processors},
  author={Choi, Mingeun and Krishnakumar, Sriharini and Khorasani, Ramin Rahimzadeh and Swaminathan, Madhavan and Partin-Vaisband, Inna and Kumar, Satish},
  journal={IEEE Transactions on Components, Packaging and Manufacturing Technology},
  year={2025},
  publisher={IEEE}
}

@article{wong2001performance,
  title={Performance improvements of interleaving VRMs with coupling inductors},
  author={Wong, Pit-Leong and Xu, Peng and Yang, P and Lee, Fred C},
  journal={IEEE Transactions on Power Electronics},
  volume={16},
  number={4},
  pages={499--507},
  year={2001},
  publisher={IEEE}
}

@article{li2022butterfly,
  title={Butterfly interleaving winding arrangements for multiphase coupled inductors},
  author={Li, Mingxiao and Liu, Yunfeng and Ouyang, Ziwei and Andersen, Michael AE},
  journal={IEEE Transactions on Power Electronics},
  volume={38},
  number={3},
  pages={3315--3327},
  year={2022},
  publisher={IEEE}
}

@article{zhang2024organic,
  title={Organic package substrate embedded coupled magnetic core inductors using lithographic via technology for power supply in package},
  author={Zhang, Weihao and Zhou, Guoyun and Hong, Yan and Chen, Xianming and Huang, Benxia and Xu, Xiaowei and Ding, Shuai and Zhu, Zhaojun and Huang, Yongmao and He, Wei},
  journal={Results in Physics},
  volume={60},
  pages={107628},
  year={2024},
  publisher={Elsevier}
}

@inproceedings{ding2020fan,
  title={Fan-out-package-embedded coupled inductors for integrated voltage conversion},
  author={Ding, Yixiao and Fang, Xiangming and Wu, Rongxiang and Sin, Johnny KO},
  booktitle={2020 32nd International Symposium on Power Semiconductor Devices and ICs (ISPSD)},
  pages={356--359},
  year={2020},
  organization={IEEE}
}

@article{wang2021novel,
  title={A novel thin film cascade matrix coupled inductor for integrated voltage regulators},
  author={Wang, Ningning and Zhou, Hui and Zhang, Zhengmin and Peng, Shanfeng and Yu, Junchao and Liao, Zhenyu and Cheng, Mengjie and Feeney, Ciaran and Liu, Lei and Ye, Tingcong},
  journal={IEEE Transactions on Power Electronics},
  volume={36},
  number={12},
  pages={13349--13354},
  year={2021},
  publisher={IEEE}
}

@article{zhu2022coupled,
  title={Coupled inductors with an adaptive coupling coefficient for multiphase voltage regulators},
  author={Zhu, Feiyang and Li, Qiang},
  journal={IEEE Transactions on Power Electronics},
  volume={38},
  number={1},
  pages={739--749},
  year={2022},
  publisher={IEEE}
}

@inproceedings{elasser2023mini22,
  title={Mini-LEGO: A 1.5-MHz 240-A 48-V-to-1-V CPU VRM with 8.4-mm height for vertical power delivery},
  author={Elasser, Youssef and Baek, Jaeil and Radhakrishnan, Kaladhar and Gan, Houle and Douglas, Jonathan and De, Vivek and Jiang, Shuai and Krishnamurthy, Harish K and Li, Xin and Sullivan, Charles R and others},
  booktitle={2023 IEEE Applied Power Electronics Conference and Exposition (APEC)},
  pages={1959--1966},
  year={2023},
  organization={IEEE}
}

@inproceedings{li2025air,
  title={Air-LEGO: A Magnetic-Free Ultra-Thin 24V-to-1V 120A VRM with Air-Coupled Inductors},
  author={Li, Haoran and Zeng, Wenliang and Elasser, Youssef and Chen, Minjie},
  booktitle={2025 IEEE Applied Power Electronics Conference and Exposition (APEC)},
  pages={510--517},
  year={2025},
  organization={IEEE}
}

@article{elasser2023mini2,
  title={Mini-LEGO CPU voltage regulator},
  author={Elasser, Youssef and Baek, Jaeil and Radhakrishnan, Kaladhar and Gan, Houle and Douglas, Jonathan P and Krishnamurthy, Harish K and Li, Xin and Jiang, Shuai and De, Vivek and Sullivan, Charles R and others},
  journal={IEEE Transactions on Power Electronics},
  volume={39},
  number={3},
  pages={3391--3410},
  year={2023},
  publisher={IEEE}
}

@INPROCEEDINGS{DT_1,
  author={Weiler, Pelle and Vermulst, Bas},
  booktitle={2020 IEEE Applied Power Electronics Conference and Exposition (APEC)}, 
  title={Gate Driver with Short Inherent Dead-Time for Wide-Bandgap High-Precision Inverters}, 
  year={2020},
  volume={},
  number={},
  pages={1593-1598},
  doi={10.1109/APEC39645.2020.9124300}}

@INPROCEEDINGS{DT_2,
  author={LaBella, Thomas and York, Ben and Hutchens, Chris and Lai, Jih-Sheng},
  booktitle={2012 IEEE Energy Conversion Congress and Exposition (ECCE)}, 
  title={Dead time optimization through loss analysis of an active-clamp flyback converter utilizing GaN devices}, 
  year={2012},
  volume={},
  number={},
  pages={3882-3889},
  doi={10.1109/ECCE.2012.6342304}}

@misc{EPC_GaNPowerBench_2025,
  author       = {{Efficient Power Conversion (EPC)}},
  title        = {{GaN FET Selection Tool for Buck Converters (GaN Power Bench)}},
  year         = {2025},
  howpublished = {\url{https://epc-co.com/epc/design-support/tools-and-calculators/gan-power-bench/gan-fet-selection-tool-for-buck-convertors}},
  note         = {Accessed: Mar. 3, 2026}
}

@inproceedings{krishnakumar2024_GLSVLSI,
  title={System Architecture Optimization for Vertical Power Delivery},
  author={Krishnakumar, Sriharini and Popryho, Yaroslav and Partin-Vaisband, Inna},
  booktitle={Proceedings of the Great Lakes Symposium on VLSI 2024},
  pages={351--352},
  year={2024}
}

@techreport{HIR_WLP_2021,
  title        = {Heterogeneous Integration Roadmap: Chapter 23 -- Wafer-Level Packaging (WLP)},
  author       = {{IEEE Electronics Packaging Society}},
  year         = {2021},
  institution  = {IEEE Electronics Packaging Society (EPS)},
  url          = {https://eps.ieee.org/wp-content/uploads/2025/11/ch23-wlpfinal2.pdf},
  note         = {Heterogeneous Integration Roadmap (HIR)}
}

@misc{EPC2040_datasheet,
  author       = {E. P. C.},
  title        = {{EPC2040 – Enhancement Mode GaN Power Transistor Datasheet}},
  year         = {2021},
  howpublished = {\url{https://epc-co.com/epc/portals/0/epc/documents/datasheets/EPC2040_datasheet.pdf}},
}

@misc{EPC2216_datasheet,
  author       = {Efficient Power Conversion Corporation},
  title        = {{EPC2216 – Enhancement Mode GaN Power Transistor Datasheet}},
  year         = {2025},
  howpublished = {\url{https://epc-co.com/epc/portals/0/epc/documents/datasheets/EPC2216_datasheet.pdf}},
}

@INPROCEEDINGS{PG_HPC,
  author={Abdelzaher, Salma and Gharib, Mohamed and Trivedi, Amit Ranjan and Partin-Vaisband, Inna},
  booktitle={2025 IEEE 34th Conference on Electrical Performance of Electronic Packaging and Systems (EPEPS)}, 
  title={Bandit Learning-Driven Power Gating with State Retention for High Performance Computing}, 
  year={2025},
  volume={},
  number={},
  pages={1-3},
  doi={10.1109/EPEPS63858.2025.11346621}}

@article{rasheedi202528,
  title={A 28 nm multiply-accumulate ASIC architecture for on-chip data compression in MHz frame rate X-ray and electron pixel detectors},
  author={Rasheedi, Rami and Contini, Nicholas and Gharib, Mohamed Adel and Strempfer, Sebastian and Gnanasekaran, Senthil and Abdelzaher, Salma and Guruswamy, Tejas and Yoshii, Kazutomo and Hammer, Mike and Shi, Henry and others},
  journal={Journal of Instrumentation},
  volume={20},
  number={10},
  pages={P10027},
  year={2025},
  publisher={IOP Publishing}
}

\begin{IEEEbiography}[{\includegraphics[width=1 in,height=1.25 in,clip,keepaspectratio]{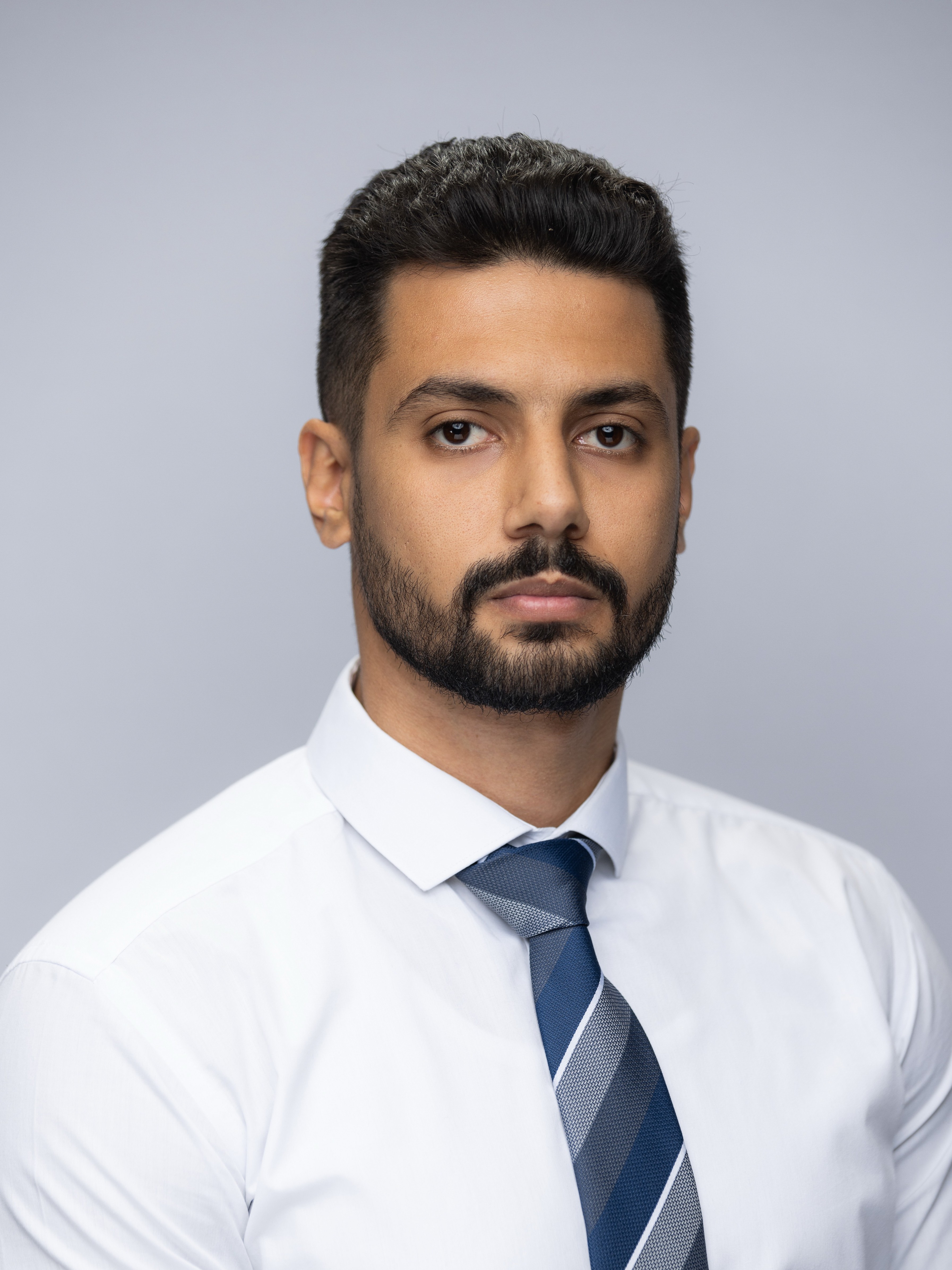}}]{Rami Rasheedi} (Graduate Student Member, IEEE) received the B.S. degree in Electronics and Communications Engineering from Ain Shams University, Cairo, Egypt, in 2021. He accumulated two years of professional experience in the semiconductor industry, and then began his Ph.D. studies at the High-Performance Circuits and Systems (HiPerCAS) Lab, University of Illinois Chicago, Chicago, IL, USA, in 2023, and has been a Visiting Student at Argonne National Laboratory, Lemont, IL, USA, since 2025. His research focuses on the modeling, design, and optimization of passive devices for vertical power delivery in chiplet-based high-performance systems.

\end{IEEEbiography}


\begin{IEEEbiography}[{\includegraphics[width=1in,height=1.25 in,clip,keepaspectratio]{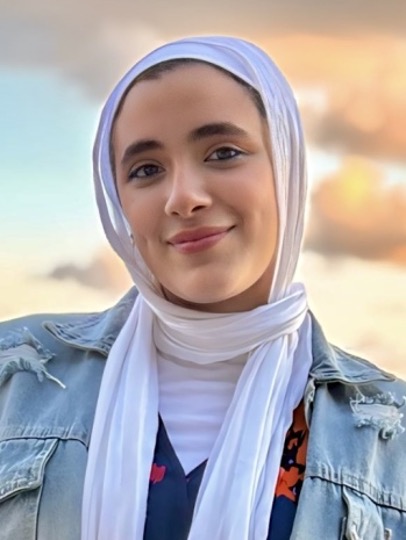}}]{Salma Abdelzaher}
(Graduate Student Member, IEEE) received the B.S. degree in Electronics and Electrical Communications Engineering from Cairo University, Cairo, Egypt, in 2021. She is currently pursuing the Ph.D. degree in Electrical and Computer Engineering at the University of Illinois Chicago, Chicago, IL, USA, where she has been a member of the High-Performance Circuits and Systems (HiPerCAS) Laboratory since 2023. Prior to her doctoral studies, she worked at Siemens EDA as an Analog QA/Test Engineer for two years. Since 2025, she has also been a research intern at Argonne National Laboratory, Lemont, IL, USA. Her research focuses on the modeling, design, and optimization of power conversion circuits for vertical power delivery in chiplet-based high-performance computing systems, as well as reinforcement-learning-based power management in heterogeneous computing platforms.
\end{IEEEbiography}

\begin{IEEEbiography}[{\includegraphics[width=1 in]{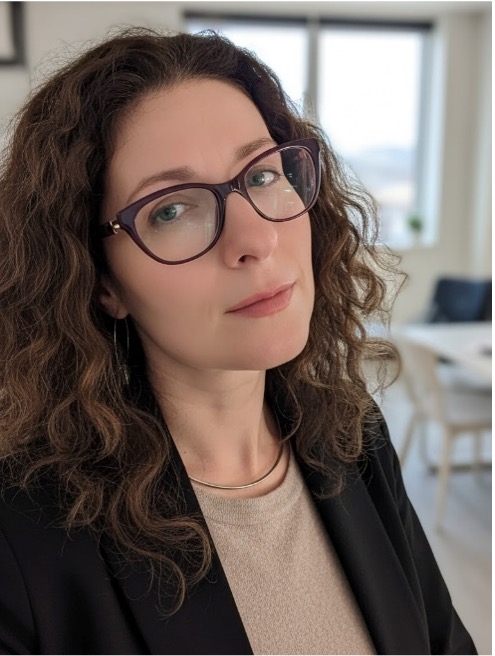}}]{Inna Partin-Vaisband}
(Senior Member, IEEE) received the Ph.D. degree in Electrical and Computer
Engineering from the University of Rochester, NY,
USA, in 2015, and the B.Sc. and M.Sc. degrees
from the Technion–Israel Institute of Technology,
Haifa, Israel, in 2006 and 2009, respectively. She
is currently an Associate Professor in the Department of Electrical and Computer Engineering at the
University of Illinois Chicago, IL, USA. Previously,
she held various software and hardware R\&D positions at several companies, including IBM, Haifa, Israel (2005–2009). Her research interests include vertical power delivery, analog
and mixed-signal circuit design, AI-assisted electronic design automation, and
hardware security. Focus is placed on 2.5D/3D edge and high-performance computing systems. She is the author of \textit{On-Chip Power
Delivery and Management} (4th ed.), and her distributed on-chip power-supply
architectures have been deployed in commercial mobile SoCs. Her recent
work on chiplet-based systems was featured in the \textit{Communications of the
ACM} article “The Chiplet Revolution” (2024). Dr. Partin-Vaisband serves as
an Associate Editor for the \textit{IEEE Transactions on Components, Packaging and Manufacturing }(CPMT) and \textit{Microelectronics
Journal}. She serves as the Editor of the \textit{IEEE T-CPMT} Special Section on Vertical Power Delivery for Next-Generation Advanced Packaging Systems. In the past, she served as the General Chair of the \textit{ACM Great Lakes Symposium
on VLSI} (GLSVLSI) 2024. She is a recipient of the 2022 Google Research
Scholar Award and the 2023 NSF CAREER Award.
\end{IEEEbiography}
\vfill

\end{document}